\pdfoutput=1
\documentclass[preprint,journal]{vgtc}       

\ifpdf 
    \pdfoutput=1\relax 
    \usepackage{graphicx} 
    \DeclareGraphicsExtensions{.pdf,.png,.jpg,.jpeg} 
\else 
    \usepackage{graphicx} 
    \DeclareGraphicsExtensions{.eps} 
\fi

\graphicspath{{figures/}{images/}{./}} 

\usepackage{microtype} 
\PassOptionsToPackage{warn}{textcomp} 
\usepackage{textcomp} 

\usepackage{mathptmx} 
\usepackage{times} 
\usepackage{cite} 
\usepackage{tabu} 
\usepackage{booktabs} 

\vgtcinsertpkg

\usepackage{amsmath}
\usepackage{gensymb}
\usepackage{amssymb}
\usepackage{enumitem}
\usepackage[usenames, dvipsnames]{color}
\usepackage[dvipsnames]{xcolor}
\usepackage[caption=false]{subfig}
\usepackage{calrsfs}
\usepackage{tabularx}
\usepackage{layouts}
\usepackage{tikz}
\usepackage{multirow}

\DeclareMathAlphabet{\pazocal}{OMS}{zplm}{m}{n}
\newcommand{\norm}[1]{\left\lVert#1\right\rVert}
\definecolor{myteal}{rgb}{0.0, 0.6, 0.6}

\onlineid{105}

\vgtccategory{Research}
\vgtcpapertype{algorithm/technique}

\title{What Would a Graph Look Like in This Layout?\\A Machine Learning Approach to Large Graph Visualization}

\author{Oh-Hyun~Kwon, \textit{Student Member,~IEEE},
Tarik~Crnovrsanin,
and Kwan-Liu~Ma, \textit{Fellow,~IEEE}
}
\authorfooter{
\item All authors are with the University of California, Davis.\protect\\E-mail: kw@ucdavis.edu, tecrnovr@ucdavis.edu, ma@cs.ucdavis.edu.
}

\shortauthortitle{Kwon \MakeLowercase{\textit{et al.}}: What Would a Graph Look Like in This Layout?}

\abstract{Using different methods for laying out a graph can lead to very different visual appearances, with which the viewer perceives different information. 
Selecting a ``good'' layout method is thus important for visualizing a graph. 
The selection can be highly subjective and dependent on the given task. 
A common approach to selecting a good layout is to use aesthetic criteria and visual inspection. 
However, fully calculating various layouts and their associated aesthetic metrics is computationally expensive. 
In this paper, we present a machine learning approach to large graph visualization based on computing the topological similarity of graphs using graph kernels. 
For a given graph, our approach can show what the graph would look like in different layouts and estimate their corresponding aesthetic metrics. 
An important contribution of our work is the development of a new framework to design graph kernels. 
Our experimental study shows that our estimation calculation is considerably faster than computing the actual layouts and their aesthetic metrics. 
Also, our graph kernels outperform the state-of-the-art ones in both time and accuracy. 
In addition, we conducted a user study to demonstrate that the topological similarity computed with our graph kernel matches perceptual similarity assessed by human users.}


\teaser{
\centering
\includegraphics[width=460pt,keepaspectratio]{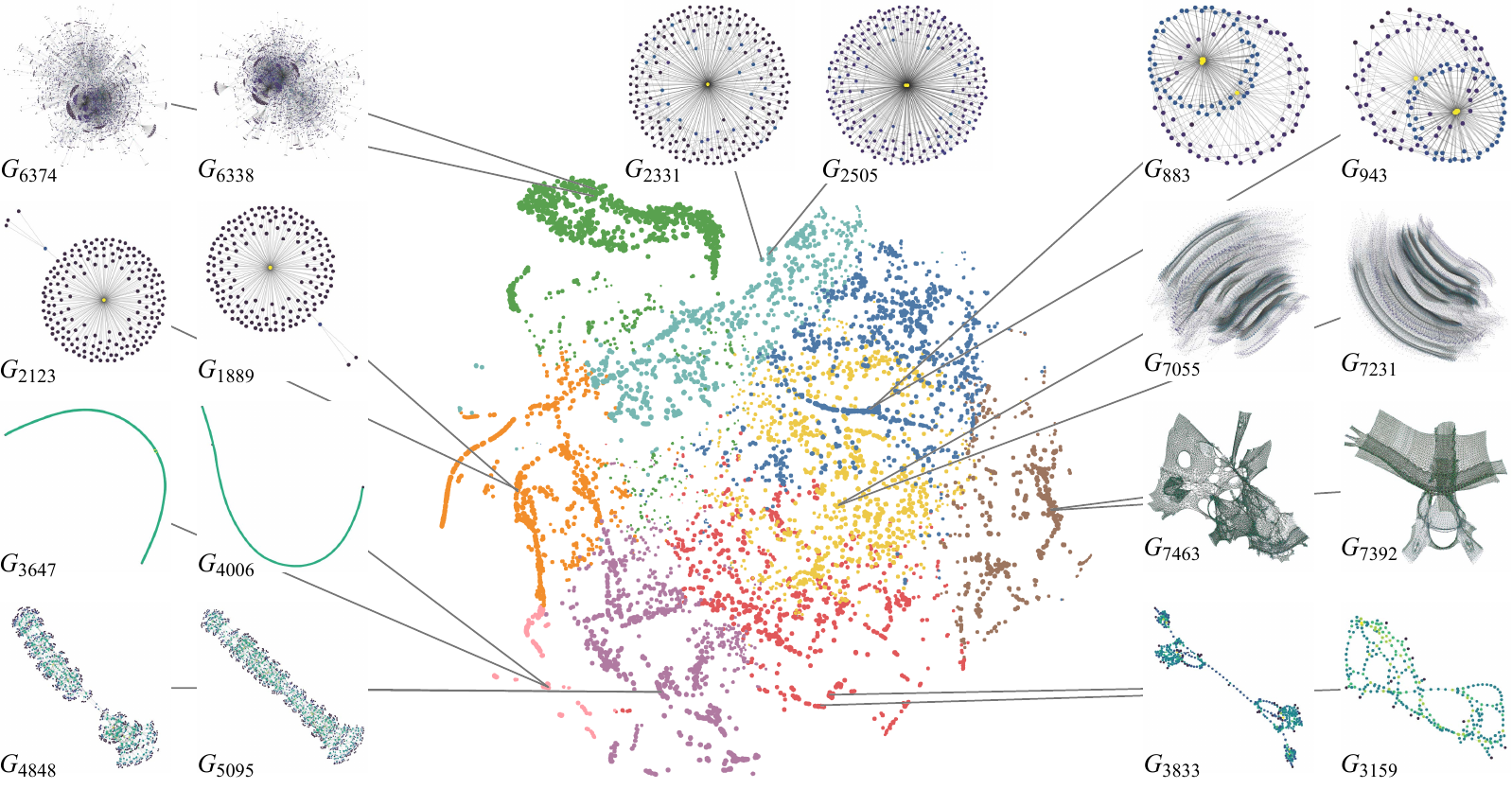}
\caption{A projection of topological similarities between 8,263 graphs measured by our \textsc{RW-Log-Laplacian} kernel. 
Based on the topological similarity, our approach shows what a graph would look like in different layouts and estimates their corresponding aesthetic metrics. 
We clustered the graphs based on their topological similarities for the purpose of the user study.
The two graphs in each pair are the most topologically similar, but not isomorphic, to each other.
The projection is computed with t-SNE \cite{TSNE}, and the highlighted graphs are visualized with FM$^3$ layout \cite{Hachul04}.
An interactive plot is available in the supplementary materials \cite{Supplementary}.}
\label{fig:teaser}
}

\begin{document}


\firstsection{Introduction}
\maketitle
Graphs are popularly used to represent complex systems, such as social networks, power grids, and biological networks.
Visualizing a graph can help us better understand the structure of the data.
Many graph visualization methods have been introduced \cite{Herman00, Landesberger11, Gibson12}, with the most popular and intuitive method being the node-link diagram.

Over the last five decades, a multitude of methods have been developed to lay out a node-link diagram. 
A graph's layout results can be greatly different depending on which layout method is used. 
Because the layout of a graph significantly influences the user's understanding of the graph \cite{McGrath96, McGrath97, Huang07, Gibson12}, it is important to find a ``good'' layout that can effectively depict the structure of the graph. 
Defining a good layout can be highly subjective and dependent on the given task. 
A suitable starting point for finding a good layout is to use both the aesthetic criteria, such as \emph{reducing edge crossings}, and the user's visual inspection.

When the graph is large, computing several graph layouts and selecting one through visual inspection and/or aesthetic metrics is, unfortunately, not a practical solution. 
The amount of time it would take to compute these various layouts and aesthetic metrics is tremendous.
For a graph with millions of vertices, a single layout can take hours or days to calculate.
In addition, we often must consider multiple aesthetic metrics to evaluate a single graph layout, since there is no consensus on which criteria are the most effective or preferable \cite{Dunne09}.
As graph data is commonly used in data analysis tasks and is expected to grow in size at even higher rates, alternative solutions are needed.

One possible solution is to quickly estimate aesthetic metrics and show what a graph would look like through predictive methods.
In the field of machine learning, several methods have been used to predict the properties of graphs, such as the classes of graphs.
One prominent approach to predicting such properties is to use a graph kernel.
Graph kernel methods enable us to apply various kernelized machine learning techniques, such as the Support Vector Machine (SVM) \cite{Cortes95}, on graphs.

In this paper, we present a machine learning approach that can show what a graph would look like in different layouts and estimate their corresponding aesthetic metrics.
The fundamental assumption of our approach is the following: given the same layout method, if the graphs have similar topological structures, then they will have similar resulting layouts.
Under this assumption, we introduce new graph kernels to measure the topological similarities between graphs.
Then, we apply machine learning techniques to show what a new input graph would look like in different layouts and estimate their corresponding aesthetic metrics.
To the best of our knowledge, this is the first time graph kernels have been utilized in the field of graph visualization.

The primary contributions of this work include:
\begin{itemize}[topsep=0pt,itemsep=-1ex,partopsep=1ex,parsep=1ex]
\item A fast and accurate method to show what a graph would look like in different layouts.
\item A fast and accurate method to estimate graph layout aesthetic metrics.
\item A framework for designing graph kernels based on graphlets.
\item A demonstration of the effectiveness of graph kernels as an approach to large graph visualization.
\end{itemize}
We evaluate our methods in two ways. 
First, we compare 13 graph kernels, which include two state-of-the-art ones, 
based on accuracy and computation time for estimating aesthetic metrics.
The results show that our estimations of aesthetic metrics are highly accurate and fast. 
Our graph kernels outperform existing kernels in both time and accuracy. 
Second, we conduct a user study to demonstrate that the topological similarity computed with our graph kernel matches the perceptual similarity assessed by human users.

\section{Background}
In this section, we present the notations and definitions used in this paper and introduce graph kernels.

\subsection{Notations and Definitions}
Let $G=(V,E)$ be a graph, where 
$V = \left\{v_1, \dots, v_n \right\}$ is a set of $n$ vertices (or nodes), 
and $E = \left\{e_1, \dots, e_m \mid e = (v_i, v_j),\ v_i, v_j \in V \right\}$ is a set of $m$ edges (or links).
An edge $e=(v_i, v_j)$ is said to be \emph{incident} to vertex $v_i$ and vertex $v_j$.
An edge that connects a vertex to itself $e=(v_i, v_i)$ is called a \emph{self loop}. 
Two or more edges that are incident to the same two vertices are called \emph{multiple edges}.
A graph is considered \emph{simple} if it contains no self loops or multiple edges.
An \emph{undirected} graph is one where $(v_i, v_j) \in E \Leftrightarrow (v_j, v_i) \in E$.
A graph is called \emph{unlabeled} if there is no distinction between vertices other than their interconnectivity.
In this paper, we consider simple, connected, undirected, and unlabeled graphs.

Given a graph $G$, a graph $G' = (V', E')$ is a \emph{subgraph} of $G$ if $V' \subseteq V$ and $E' \subseteq E$.
A subgraph $G' = (V', E')$ is called an \emph{induced (or vertex-induced)} subgraph of $G$ if $E' = \{(v_i, v_j) \mid (v_i, v_j) \in E \text{ and } v_i, v_j \in V' \}$, that is, all edges in $E$, between two vertices $v_i, v_j \in V'$, are also present in $E'$.
Two graphs $G=(V,E)$ and $G'=(V',E')$ are \emph{isomorphic} if there exists a bijection $f: V \rightarrow V'$, called \emph{isomorphism}, such that $(v_i, v_j) \in E \Leftrightarrow (f(v_i), f(v_j)) \in E'$ for all $v_i, v_j \in V$.

Suppose we have empirical data $(x_1, y_1), \dots, (x_n, y_n) \in \pazocal{X} \times \pazocal{Y}$, where the domain $\pazocal{X}$ is a nonempty set of \emph{inputs} $x_i$ and $\pazocal{Y}$ is a set of corresponding \emph{targets} $y_i$.
A kernel method predicts the target $y$ of a new input $x$ based on existing ``similar'' inputs and their outputs $(x_i, y_i)$.
A function $k: \pazocal{X} \times \pazocal{X}  \mapsto \mathbb{R}$ that measures the similarity between two inputs is called a kernel function, or simply a kernel.
A kernel function $k$ is often defined as an inner product of two vectors in a \emph{feature space} $\pazocal{H}$:
\vspace{-0.25em}
\begin{equation*}
    k(x, x') = \left\langle \phi(x), \phi(x') \right\rangle = \left\langle \mathbf{x}, \mathbf{x'} \right\rangle
\end{equation*}
where $\phi: \pazocal{X} \mapsto \pazocal{H}$ is called a \emph{feature map} which maps an input $x$ to a \emph{feature vector} $\mathbf{x}$ in  $\pazocal{H}$.

\subsection{Measuring Topological Similarities between Graphs}
Based on our assumption, we need to measure the topological similarities between graphs.
Depending on the discipline, this problem is called \textit{graph matching}, \textit{graph comparison}, or \textit{network alignment}.
In the last five decades, numerous approaches have been proposed to this problem \cite{Dehmer14, EmmertStreib16}.
Each approach measures the similarity between graphs based on different aspects, such as isomorphism relation \cite{Zelinka75, Sobik82, Kaden82}, graph edit distance \cite{Bunke83}, or graph measures \cite{Costa07, Barabasi16}.
However, many of these traditional approaches are either computationally expensive, not expressive enough to capture the topological features, or difficult to adapt to different problems \cite{Shervashidze12}.

Graph kernels have been recently introduced for measuring pairwise similarities between graphs in the field of machine learning.
They allow us to apply many different kernelized machine learning techniques, e.g. SVM \cite{Cortes95}, on graph problems, including graph classification problems found in bioinformatics and chemoinformatics.

A graph kernel can be considered to be an instance of an R-convolution kernel \cite{Haussler99}.
It measures the similarity between two graphs based on the recursively decomposed substructures of said graphs.
The measure of similarity varies with each graph kernel based on different types of substructures in graphs.
These substructures include walks \cite{Gartner03, Kashima04, Vishwanathan10}, shortest paths \cite{Borgwardt05}, subtrees \cite{Ramon03, Shervashidze09B, Shervashidze11}, cycles \cite{Horvath04}, and graphlets \cite{Shervashidze09}.
Selecting a graph kernel is challenging as many kernels are available.
To exacerbate the problem, there is no theoretical justification as to why one graph kernel works better than another for a given problem \cite{Shervashidze09}.

Many graph kernels often have similar limitations as previously mentioned approaches.
They do not scale well to large graphs (time complexity of $O(|V|^3)$ or higher) or do not work well for unlabeled graphs.
To overcome this problem, a graph kernel based on sampling a fixed number of graphlets has been introduced to be accurate and efficient on large graphs \cite{Shervashidze09, Shervashidze12}.

\begin{figure}[tb]
\centering
\includegraphics[width=\columnwidth, keepaspectratio]{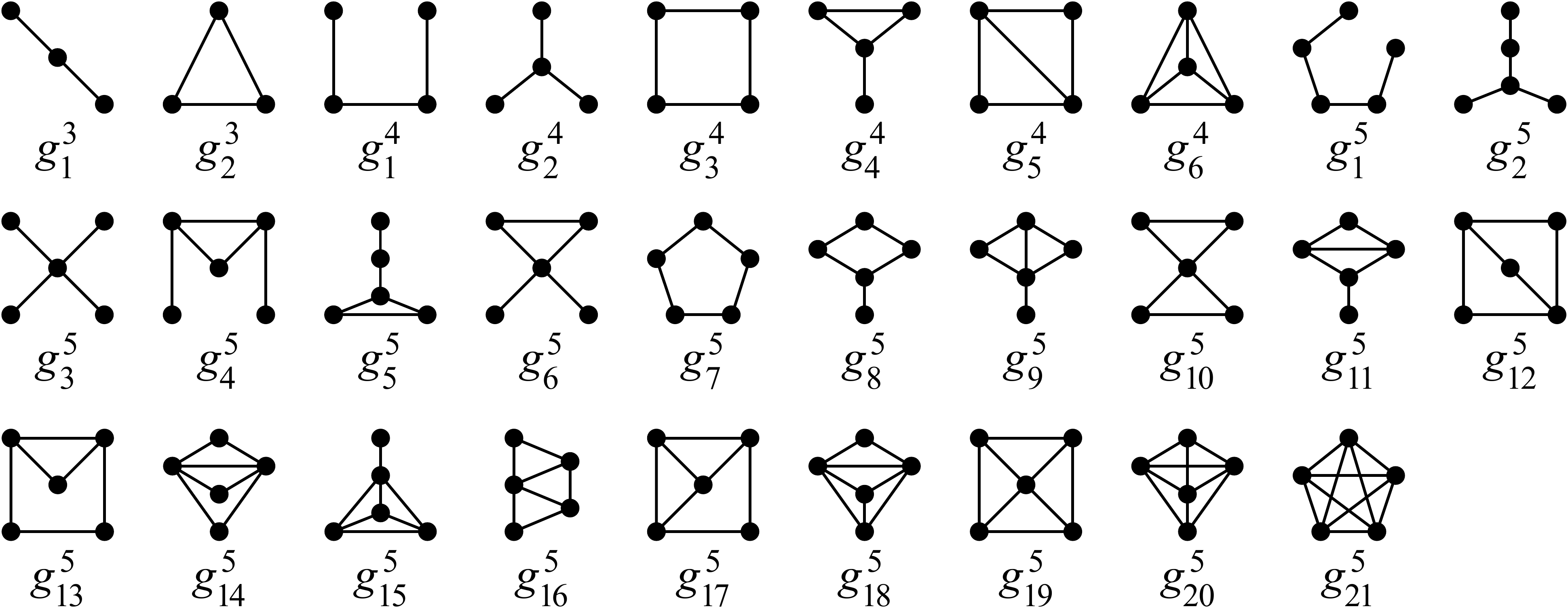}
\vspace{-1.5em}
\caption{All connected graphlets of 3, 4, or 5 vertices.}
\vspace{-.7em}
\label{fig:graphlets}
\end{figure}

\begin{figure*}[htb!]
\captionsetup{farskip=0pt}
\centering
\subfloat[$G_{883}\quad|V|=122\quad|E|=472$]{\label{fig:graphlet-frequency-a}\includegraphics[width=.33\columnwidth,keepaspectratio]{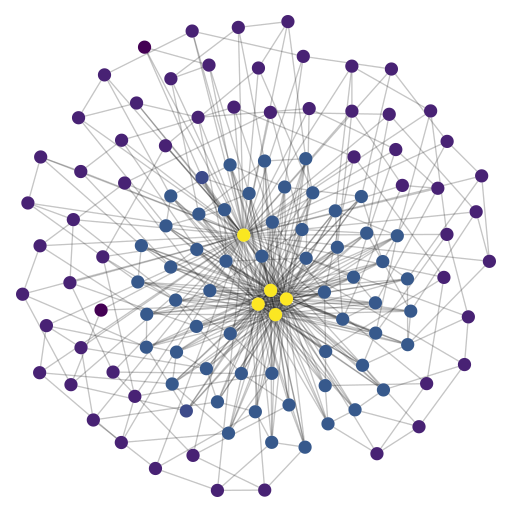}
\begin{tikzpicture}[x=1pt,y=1pt]
\definecolor{fillColor}{RGB}{255,255,255}
\path[use as bounding box,fill=fillColor,fill opacity=0.00] (0,0) rectangle (165.86, 82.89);
\begin{scope}
\path[clip] (  7.95, 12.83) rectangle (165.86, 82.89);
\definecolor{drawColor}{gray}{0.92}

\path[draw=drawColor,line width= 0.2pt,line join=round] (  7.95, 25.12) --
	(165.86, 25.12);

\path[draw=drawColor,line width= 0.2pt,line join=round] (  7.95, 43.31) --
	(165.86, 43.31);

\path[draw=drawColor,line width= 0.2pt,line join=round] (  7.95, 61.51) --
	(165.86, 61.51);

\path[draw=drawColor,line width= 0.2pt,line join=round] (  7.95, 79.71) --
	(165.86, 79.71);

\path[draw=drawColor,line width= 0.2pt,line join=round] (  7.95, 16.02) --
	(165.86, 16.02);

\path[draw=drawColor,line width= 0.2pt,line join=round] (  7.95, 34.21) --
	(165.86, 34.21);

\path[draw=drawColor,line width= 0.2pt,line join=round] (  7.95, 52.41) --
	(165.86, 52.41);

\path[draw=drawColor,line width= 0.2pt,line join=round] (  7.95, 70.61) --
	(165.86, 70.61);

\path[draw=drawColor,line width= 0.2pt,line join=round] ( 11.19, 12.83) --
	( 11.19, 82.89);

\path[draw=drawColor,line width= 0.2pt,line join=round] ( 16.60, 12.83) --
	( 16.60, 82.89);

\path[draw=drawColor,line width= 0.2pt,line join=round] ( 22.01, 12.83) --
	( 22.01, 82.89);

\path[draw=drawColor,line width= 0.2pt,line join=round] ( 27.42, 12.83) --
	( 27.42, 82.89);

\path[draw=drawColor,line width= 0.2pt,line join=round] ( 32.83, 12.83) --
	( 32.83, 82.89);

\path[draw=drawColor,line width= 0.2pt,line join=round] ( 38.23, 12.83) --
	( 38.23, 82.89);

\path[draw=drawColor,line width= 0.2pt,line join=round] ( 43.64, 12.83) --
	( 43.64, 82.89);

\path[draw=drawColor,line width= 0.2pt,line join=round] ( 49.05, 12.83) --
	( 49.05, 82.89);

\path[draw=drawColor,line width= 0.2pt,line join=round] ( 54.46, 12.83) --
	( 54.46, 82.89);

\path[draw=drawColor,line width= 0.2pt,line join=round] ( 59.87, 12.83) --
	( 59.87, 82.89);

\path[draw=drawColor,line width= 0.2pt,line join=round] ( 65.27, 12.83) --
	( 65.27, 82.89);

\path[draw=drawColor,line width= 0.2pt,line join=round] ( 70.68, 12.83) --
	( 70.68, 82.89);

\path[draw=drawColor,line width= 0.2pt,line join=round] ( 76.09, 12.83) --
	( 76.09, 82.89);

\path[draw=drawColor,line width= 0.2pt,line join=round] ( 81.50, 12.83) --
	( 81.50, 82.89);

\path[draw=drawColor,line width= 0.2pt,line join=round] ( 86.90, 12.83) --
	( 86.90, 82.89);

\path[draw=drawColor,line width= 0.2pt,line join=round] ( 92.31, 12.83) --
	( 92.31, 82.89);

\path[draw=drawColor,line width= 0.2pt,line join=round] ( 97.72, 12.83) --
	( 97.72, 82.89);

\path[draw=drawColor,line width= 0.2pt,line join=round] (103.13, 12.83) --
	(103.13, 82.89);

\path[draw=drawColor,line width= 0.2pt,line join=round] (108.54, 12.83) --
	(108.54, 82.89);

\path[draw=drawColor,line width= 0.2pt,line join=round] (113.94, 12.83) --
	(113.94, 82.89);

\path[draw=drawColor,line width= 0.2pt,line join=round] (119.35, 12.83) --
	(119.35, 82.89);

\path[draw=drawColor,line width= 0.2pt,line join=round] (124.76, 12.83) --
	(124.76, 82.89);

\path[draw=drawColor,line width= 0.2pt,line join=round] (130.17, 12.83) --
	(130.17, 82.89);

\path[draw=drawColor,line width= 0.2pt,line join=round] (135.58, 12.83) --
	(135.58, 82.89);

\path[draw=drawColor,line width= 0.2pt,line join=round] (140.98, 12.83) --
	(140.98, 82.89);

\path[draw=drawColor,line width= 0.2pt,line join=round] (146.39, 12.83) --
	(146.39, 82.89);

\path[draw=drawColor,line width= 0.2pt,line join=round] (151.80, 12.83) --
	(151.80, 82.89);

\path[draw=drawColor,line width= 0.2pt,line join=round] (157.21, 12.83) --
	(157.21, 82.89);

\path[draw=drawColor,line width= 0.2pt,line join=round] (162.61, 12.83) --
	(162.61, 82.89);
\definecolor{fillColor}{gray}{0.35}

\path[fill=fillColor] (  9.84, 16.02) rectangle ( 12.55, 55.33);

\path[fill=fillColor] ( 15.25, 16.02) rectangle ( 17.95, 45.13);

\path[fill=fillColor] ( 20.66, 16.02) rectangle ( 23.36, 59.07);

\path[fill=fillColor] ( 26.07, 16.02) rectangle ( 28.77, 66.48);

\path[fill=fillColor] ( 31.47, 16.02) rectangle ( 34.18, 37.95);

\path[fill=fillColor] ( 36.88, 16.02) rectangle ( 39.59, 58.25);

\path[fill=fillColor] ( 42.29, 16.02) rectangle ( 44.99, 56.08);

\path[fill=fillColor] ( 47.70, 16.02) rectangle ( 50.40, 45.10);

\path[fill=fillColor] ( 53.11, 16.02) rectangle ( 55.81, 62.91);

\path[fill=fillColor] ( 58.51, 16.02) rectangle ( 61.22, 71.81);

\path[fill=fillColor] ( 63.92, 16.02) rectangle ( 66.63, 76.29);

\path[fill=fillColor] ( 69.33, 16.02) rectangle ( 72.03, 60.02);

\path[fill=fillColor] ( 74.74, 16.02) rectangle ( 77.44, 61.14);

\path[fill=fillColor] ( 80.15, 16.02) rectangle ( 82.85, 70.36);

\path[fill=fillColor] ( 85.55, 16.02) rectangle ( 88.26, 34.79);

\path[fill=fillColor] ( 90.96, 16.02) rectangle ( 93.66, 52.93);

\path[fill=fillColor] ( 96.37, 16.02) rectangle ( 99.07, 59.48);

\path[fill=fillColor] (101.78, 16.02) rectangle (104.48, 54.95);

\path[fill=fillColor] (107.18, 16.02) rectangle (109.89, 58.48);

\path[fill=fillColor] (112.59, 16.02) rectangle (115.30, 16.02);

\path[fill=fillColor] (118.00, 16.02) rectangle (120.70, 45.25);

\path[fill=fillColor] (123.41, 16.02) rectangle (126.11, 66.96);

\path[fill=fillColor] (128.82, 16.02) rectangle (131.52, 49.18);

\path[fill=fillColor] (134.22, 16.02) rectangle (136.93, 41.39);

\path[fill=fillColor] (139.63, 16.02) rectangle (142.34, 16.02);

\path[fill=fillColor] (145.04, 16.02) rectangle (147.74, 58.33);

\path[fill=fillColor] (150.45, 16.02) rectangle (153.15, 16.02);

\path[fill=fillColor] (155.86, 16.02) rectangle (158.56, 54.70);

\path[fill=fillColor] (161.26, 16.02) rectangle (163.97, 42.07);
\end{scope}
\begin{scope}
\path[clip] (  0.00,  0.00) rectangle (165.86, 82.89);
\definecolor{drawColor}{gray}{0.30}

\node[text=drawColor,anchor=base east,inner sep=0pt, outer sep=0pt, scale=  0.60] at (  3.00, 14.06) {0};

\node[text=drawColor,anchor=base east,inner sep=0pt, outer sep=0pt, scale=  0.60] at (  3.00, 32.26) {2};

\node[text=drawColor,anchor=base east,inner sep=0pt, outer sep=0pt, scale=  0.60] at (  3.00, 50.46) {4};

\node[text=drawColor,anchor=base east,inner sep=0pt, outer sep=0pt, scale=  0.60] at (  3.00, 68.66) {6};
\end{scope}
\begin{scope}
\path[clip] (  0.00,  0.00) rectangle (165.86, 82.89);
\definecolor{drawColor}{gray}{0.30}

\node[text=drawColor,anchor=base,inner sep=0pt, outer sep=0pt, scale=  0.75] at ( 11.19,  3.00) {$g^3_1$};

\node[text=drawColor,anchor=base,inner sep=0pt, outer sep=0pt, scale=  0.75] at ( 22.01,  3.00) {$g^4_1$};

\node[text=drawColor,anchor=base,inner sep=0pt, outer sep=0pt, scale=  0.75] at ( 32.83,  3.00) {$g^4_3$};

\node[text=drawColor,anchor=base,inner sep=0pt, outer sep=0pt, scale=  0.75] at ( 43.64,  3.00) {$g^4_5$};

\node[text=drawColor,anchor=base,inner sep=0pt, outer sep=0pt, scale=  0.75] at ( 54.46,  3.00) {$g^5_1$};

\node[text=drawColor,anchor=base,inner sep=0pt, outer sep=0pt, scale=  0.75] at ( 65.27,  3.00) {$g^5_3$};

\node[text=drawColor,anchor=base,inner sep=0pt, outer sep=0pt, scale=  0.75] at ( 76.09,  3.00) {$g^5_5$};

\node[text=drawColor,anchor=base,inner sep=0pt, outer sep=0pt, scale=  0.75] at ( 86.90,  3.00) {$g^5_7$};

\node[text=drawColor,anchor=base,inner sep=0pt, outer sep=0pt, scale=  0.75] at ( 97.72,  3.00) {$g^5_9$};

\node[text=drawColor,anchor=base,inner sep=0pt, outer sep=0pt, scale=  0.75] at (108.54,  3.00) {$g^5_{11}$};

\node[text=drawColor,anchor=base,inner sep=0pt, outer sep=0pt, scale=  0.75] at (119.35,  3.00) {$g^5_{13}$};

\node[text=drawColor,anchor=base,inner sep=0pt, outer sep=0pt, scale=  0.75] at (130.17,  3.00) {$g^5_{15}$};

\node[text=drawColor,anchor=base,inner sep=0pt, outer sep=0pt, scale=  0.75] at (140.98,  3.00) {$g^5_{17}$};

\node[text=drawColor,anchor=base,inner sep=0pt, outer sep=0pt, scale=  0.75] at (151.80,  3.00) {$g^5_{19}$};

\node[text=drawColor,anchor=base,inner sep=0pt, outer sep=0pt, scale=  0.75] at (162.61,  3.00) {$g^5_{21}$};
\end{scope}
\end{tikzpicture}}\hfill
\subfloat[$G_{1788}\quad|V|=158\quad|E|=312$]{\label{fig:graphlet-frequency-b}\includegraphics[width=.33\columnwidth,keepaspectratio]{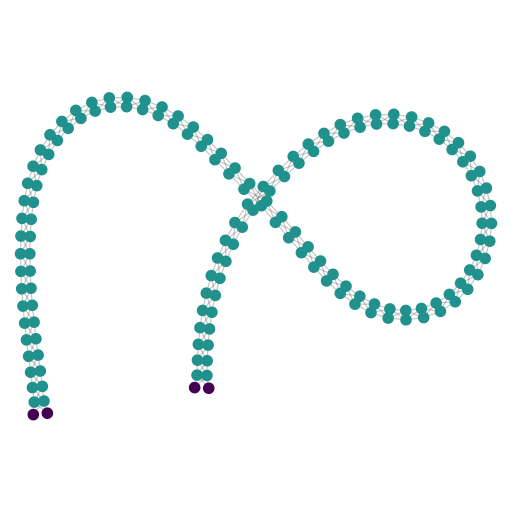}
\begin{tikzpicture}[x=1pt,y=1pt]
\definecolor{fillColor}{RGB}{255,255,255}
\path[use as bounding box,fill=fillColor,fill opacity=0.00] (0,0) rectangle (165.86, 82.89);
\begin{scope}
\path[clip] (  7.95, 12.83) rectangle (165.86, 82.89);
\definecolor{drawColor}{gray}{0.92}

\path[draw=drawColor,line width= 0.2pt,line join=round] (  7.95, 25.12) --
	(165.86, 25.12);

\path[draw=drawColor,line width= 0.2pt,line join=round] (  7.95, 43.31) --
	(165.86, 43.31);

\path[draw=drawColor,line width= 0.2pt,line join=round] (  7.95, 61.51) --
	(165.86, 61.51);

\path[draw=drawColor,line width= 0.2pt,line join=round] (  7.95, 79.71) --
	(165.86, 79.71);

\path[draw=drawColor,line width= 0.2pt,line join=round] (  7.95, 16.02) --
	(165.86, 16.02);

\path[draw=drawColor,line width= 0.2pt,line join=round] (  7.95, 34.21) --
	(165.86, 34.21);

\path[draw=drawColor,line width= 0.2pt,line join=round] (  7.95, 52.41) --
	(165.86, 52.41);

\path[draw=drawColor,line width= 0.2pt,line join=round] (  7.95, 70.61) --
	(165.86, 70.61);

\path[draw=drawColor,line width= 0.2pt,line join=round] ( 11.19, 12.83) --
	( 11.19, 82.89);

\path[draw=drawColor,line width= 0.2pt,line join=round] ( 16.60, 12.83) --
	( 16.60, 82.89);

\path[draw=drawColor,line width= 0.2pt,line join=round] ( 22.01, 12.83) --
	( 22.01, 82.89);

\path[draw=drawColor,line width= 0.2pt,line join=round] ( 27.42, 12.83) --
	( 27.42, 82.89);

\path[draw=drawColor,line width= 0.2pt,line join=round] ( 32.83, 12.83) --
	( 32.83, 82.89);

\path[draw=drawColor,line width= 0.2pt,line join=round] ( 38.23, 12.83) --
	( 38.23, 82.89);

\path[draw=drawColor,line width= 0.2pt,line join=round] ( 43.64, 12.83) --
	( 43.64, 82.89);

\path[draw=drawColor,line width= 0.2pt,line join=round] ( 49.05, 12.83) --
	( 49.05, 82.89);

\path[draw=drawColor,line width= 0.2pt,line join=round] ( 54.46, 12.83) --
	( 54.46, 82.89);

\path[draw=drawColor,line width= 0.2pt,line join=round] ( 59.87, 12.83) --
	( 59.87, 82.89);

\path[draw=drawColor,line width= 0.2pt,line join=round] ( 65.27, 12.83) --
	( 65.27, 82.89);

\path[draw=drawColor,line width= 0.2pt,line join=round] ( 70.68, 12.83) --
	( 70.68, 82.89);

\path[draw=drawColor,line width= 0.2pt,line join=round] ( 76.09, 12.83) --
	( 76.09, 82.89);

\path[draw=drawColor,line width= 0.2pt,line join=round] ( 81.50, 12.83) --
	( 81.50, 82.89);

\path[draw=drawColor,line width= 0.2pt,line join=round] ( 86.90, 12.83) --
	( 86.90, 82.89);

\path[draw=drawColor,line width= 0.2pt,line join=round] ( 92.31, 12.83) --
	( 92.31, 82.89);

\path[draw=drawColor,line width= 0.2pt,line join=round] ( 97.72, 12.83) --
	( 97.72, 82.89);

\path[draw=drawColor,line width= 0.2pt,line join=round] (103.13, 12.83) --
	(103.13, 82.89);

\path[draw=drawColor,line width= 0.2pt,line join=round] (108.54, 12.83) --
	(108.54, 82.89);

\path[draw=drawColor,line width= 0.2pt,line join=round] (113.94, 12.83) --
	(113.94, 82.89);

\path[draw=drawColor,line width= 0.2pt,line join=round] (119.35, 12.83) --
	(119.35, 82.89);

\path[draw=drawColor,line width= 0.2pt,line join=round] (124.76, 12.83) --
	(124.76, 82.89);

\path[draw=drawColor,line width= 0.2pt,line join=round] (130.17, 12.83) --
	(130.17, 82.89);

\path[draw=drawColor,line width= 0.2pt,line join=round] (135.58, 12.83) --
	(135.58, 82.89);

\path[draw=drawColor,line width= 0.2pt,line join=round] (140.98, 12.83) --
	(140.98, 82.89);

\path[draw=drawColor,line width= 0.2pt,line join=round] (146.39, 12.83) --
	(146.39, 82.89);

\path[draw=drawColor,line width= 0.2pt,line join=round] (151.80, 12.83) --
	(151.80, 82.89);

\path[draw=drawColor,line width= 0.2pt,line join=round] (157.21, 12.83) --
	(157.21, 82.89);

\path[draw=drawColor,line width= 0.2pt,line join=round] (162.61, 12.83) --
	(162.61, 82.89);
\definecolor{fillColor}{gray}{0.35}

\path[fill=fillColor] (  9.84, 16.02) rectangle ( 12.55, 56.74);

\path[fill=fillColor] ( 15.25, 16.02) rectangle ( 17.95, 16.02);

\path[fill=fillColor] ( 20.66, 16.02) rectangle ( 23.36, 58.23);

\path[fill=fillColor] ( 26.07, 16.02) rectangle ( 28.77, 56.04);

\path[fill=fillColor] ( 31.47, 16.02) rectangle ( 34.18, 54.16);

\path[fill=fillColor] ( 36.88, 16.02) rectangle ( 39.59, 16.02);

\path[fill=fillColor] ( 42.29, 16.02) rectangle ( 44.99, 16.02);

\path[fill=fillColor] ( 47.70, 16.02) rectangle ( 50.40, 16.02);

\path[fill=fillColor] ( 53.11, 16.02) rectangle ( 55.81, 59.73);

\path[fill=fillColor] ( 58.51, 16.02) rectangle ( 61.22, 57.13);

\path[fill=fillColor] ( 63.92, 16.02) rectangle ( 66.63, 48.99);

\path[fill=fillColor] ( 69.33, 16.02) rectangle ( 72.03, 16.02);

\path[fill=fillColor] ( 74.74, 16.02) rectangle ( 77.44, 16.02);

\path[fill=fillColor] ( 80.15, 16.02) rectangle ( 82.85, 16.02);

\path[fill=fillColor] ( 85.55, 16.02) rectangle ( 88.26, 16.02);

\path[fill=fillColor] ( 90.96, 16.02) rectangle ( 93.66, 57.10);

\path[fill=fillColor] ( 96.37, 16.02) rectangle ( 99.07, 16.02);

\path[fill=fillColor] (101.78, 16.02) rectangle (104.48, 16.02);

\path[fill=fillColor] (107.18, 16.02) rectangle (109.89, 16.02);

\path[fill=fillColor] (112.59, 16.02) rectangle (115.30, 51.67);

\path[fill=fillColor] (118.00, 16.02) rectangle (120.70, 16.02);

\path[fill=fillColor] (123.41, 16.02) rectangle (126.11, 16.02);

\path[fill=fillColor] (128.82, 16.02) rectangle (131.52, 16.02);

\path[fill=fillColor] (134.22, 16.02) rectangle (136.93, 16.02);

\path[fill=fillColor] (139.63, 16.02) rectangle (142.34, 16.02);

\path[fill=fillColor] (145.04, 16.02) rectangle (147.74, 16.02);

\path[fill=fillColor] (150.45, 16.02) rectangle (153.15, 16.02);

\path[fill=fillColor] (155.86, 16.02) rectangle (158.56, 16.02);

\path[fill=fillColor] (161.26, 16.02) rectangle (163.97, 16.02);
\end{scope}
\begin{scope}
\path[clip] (  0.00,  0.00) rectangle (165.86, 82.89);
\definecolor{drawColor}{gray}{0.30}

\node[text=drawColor,anchor=base east,inner sep=0pt, outer sep=0pt, scale=  0.60] at (  3.00, 14.06) {0};

\node[text=drawColor,anchor=base east,inner sep=0pt, outer sep=0pt, scale=  0.60] at (  3.00, 32.26) {2};

\node[text=drawColor,anchor=base east,inner sep=0pt, outer sep=0pt, scale=  0.60] at (  3.00, 50.46) {4};

\node[text=drawColor,anchor=base east,inner sep=0pt, outer sep=0pt, scale=  0.60] at (  3.00, 68.66) {6};
\end{scope}
\begin{scope}
\path[clip] (  0.00,  0.00) rectangle (165.86, 82.89);
\definecolor{drawColor}{gray}{0.30}

\node[text=drawColor,anchor=base,inner sep=0pt, outer sep=0pt, scale=  0.75] at ( 11.19,  3.00) {$g^3_1$};

\node[text=drawColor,anchor=base,inner sep=0pt, outer sep=0pt, scale=  0.75] at ( 22.01,  3.00) {$g^4_1$};

\node[text=drawColor,anchor=base,inner sep=0pt, outer sep=0pt, scale=  0.75] at ( 32.83,  3.00) {$g^4_3$};

\node[text=drawColor,anchor=base,inner sep=0pt, outer sep=0pt, scale=  0.75] at ( 43.64,  3.00) {$g^4_5$};

\node[text=drawColor,anchor=base,inner sep=0pt, outer sep=0pt, scale=  0.75] at ( 54.46,  3.00) {$g^5_1$};

\node[text=drawColor,anchor=base,inner sep=0pt, outer sep=0pt, scale=  0.75] at ( 65.27,  3.00) {$g^5_3$};

\node[text=drawColor,anchor=base,inner sep=0pt, outer sep=0pt, scale=  0.75] at ( 76.09,  3.00) {$g^5_5$};

\node[text=drawColor,anchor=base,inner sep=0pt, outer sep=0pt, scale=  0.75] at ( 86.90,  3.00) {$g^5_7$};

\node[text=drawColor,anchor=base,inner sep=0pt, outer sep=0pt, scale=  0.75] at ( 97.72,  3.00) {$g^5_9$};

\node[text=drawColor,anchor=base,inner sep=0pt, outer sep=0pt, scale=  0.75] at (108.54,  3.00) {$g^5_{11}$};

\node[text=drawColor,anchor=base,inner sep=0pt, outer sep=0pt, scale=  0.75] at (119.35,  3.00) {$g^5_{13}$};

\node[text=drawColor,anchor=base,inner sep=0pt, outer sep=0pt, scale=  0.75] at (130.17,  3.00) {$g^5_{15}$};

\node[text=drawColor,anchor=base,inner sep=0pt, outer sep=0pt, scale=  0.75] at (140.98,  3.00) {$g^5_{17}$};

\node[text=drawColor,anchor=base,inner sep=0pt, outer sep=0pt, scale=  0.75] at (151.80,  3.00) {$g^5_{19}$};

\node[text=drawColor,anchor=base,inner sep=0pt, outer sep=0pt, scale=  0.75] at (162.61,  3.00) {$g^5_{21}$};
\end{scope}
\end{tikzpicture}}\\
\subfloat[$G_{943}\quad|V|=124\quad|E|=462$]{\label{fig:graphlet-frequency-c}\includegraphics[width=.33\columnwidth,keepaspectratio]{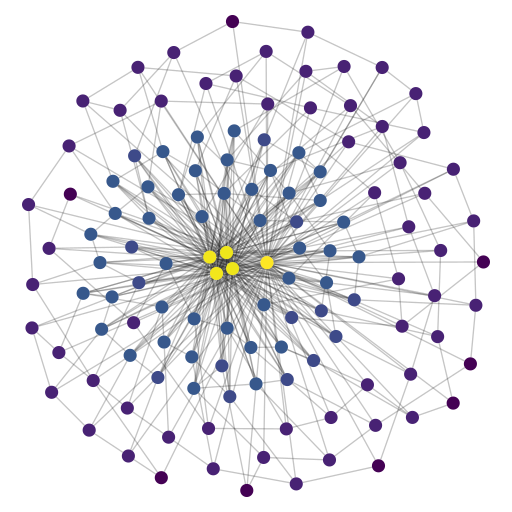}
\begin{tikzpicture}[x=1pt,y=1pt]
\definecolor{fillColor}{RGB}{255,255,255}
\path[use as bounding box,fill=fillColor,fill opacity=0.00] (0,0) rectangle (165.86, 82.89);
\begin{scope}
\path[clip] (  7.95, 12.83) rectangle (165.86, 82.89);
\definecolor{drawColor}{gray}{0.92}

\path[draw=drawColor,line width= 0.2pt,line join=round] (  7.95, 25.12) --
	(165.86, 25.12);

\path[draw=drawColor,line width= 0.2pt,line join=round] (  7.95, 43.31) --
	(165.86, 43.31);

\path[draw=drawColor,line width= 0.2pt,line join=round] (  7.95, 61.51) --
	(165.86, 61.51);

\path[draw=drawColor,line width= 0.2pt,line join=round] (  7.95, 79.71) --
	(165.86, 79.71);

\path[draw=drawColor,line width= 0.2pt,line join=round] (  7.95, 16.02) --
	(165.86, 16.02);

\path[draw=drawColor,line width= 0.2pt,line join=round] (  7.95, 34.21) --
	(165.86, 34.21);

\path[draw=drawColor,line width= 0.2pt,line join=round] (  7.95, 52.41) --
	(165.86, 52.41);

\path[draw=drawColor,line width= 0.2pt,line join=round] (  7.95, 70.61) --
	(165.86, 70.61);

\path[draw=drawColor,line width= 0.2pt,line join=round] ( 11.19, 12.83) --
	( 11.19, 82.89);

\path[draw=drawColor,line width= 0.2pt,line join=round] ( 16.60, 12.83) --
	( 16.60, 82.89);

\path[draw=drawColor,line width= 0.2pt,line join=round] ( 22.01, 12.83) --
	( 22.01, 82.89);

\path[draw=drawColor,line width= 0.2pt,line join=round] ( 27.42, 12.83) --
	( 27.42, 82.89);

\path[draw=drawColor,line width= 0.2pt,line join=round] ( 32.83, 12.83) --
	( 32.83, 82.89);

\path[draw=drawColor,line width= 0.2pt,line join=round] ( 38.23, 12.83) --
	( 38.23, 82.89);

\path[draw=drawColor,line width= 0.2pt,line join=round] ( 43.64, 12.83) --
	( 43.64, 82.89);

\path[draw=drawColor,line width= 0.2pt,line join=round] ( 49.05, 12.83) --
	( 49.05, 82.89);

\path[draw=drawColor,line width= 0.2pt,line join=round] ( 54.46, 12.83) --
	( 54.46, 82.89);

\path[draw=drawColor,line width= 0.2pt,line join=round] ( 59.87, 12.83) --
	( 59.87, 82.89);

\path[draw=drawColor,line width= 0.2pt,line join=round] ( 65.27, 12.83) --
	( 65.27, 82.89);

\path[draw=drawColor,line width= 0.2pt,line join=round] ( 70.68, 12.83) --
	( 70.68, 82.89);

\path[draw=drawColor,line width= 0.2pt,line join=round] ( 76.09, 12.83) --
	( 76.09, 82.89);

\path[draw=drawColor,line width= 0.2pt,line join=round] ( 81.50, 12.83) --
	( 81.50, 82.89);

\path[draw=drawColor,line width= 0.2pt,line join=round] ( 86.90, 12.83) --
	( 86.90, 82.89);

\path[draw=drawColor,line width= 0.2pt,line join=round] ( 92.31, 12.83) --
	( 92.31, 82.89);

\path[draw=drawColor,line width= 0.2pt,line join=round] ( 97.72, 12.83) --
	( 97.72, 82.89);

\path[draw=drawColor,line width= 0.2pt,line join=round] (103.13, 12.83) --
	(103.13, 82.89);

\path[draw=drawColor,line width= 0.2pt,line join=round] (108.54, 12.83) --
	(108.54, 82.89);

\path[draw=drawColor,line width= 0.2pt,line join=round] (113.94, 12.83) --
	(113.94, 82.89);

\path[draw=drawColor,line width= 0.2pt,line join=round] (119.35, 12.83) --
	(119.35, 82.89);

\path[draw=drawColor,line width= 0.2pt,line join=round] (124.76, 12.83) --
	(124.76, 82.89);

\path[draw=drawColor,line width= 0.2pt,line join=round] (130.17, 12.83) --
	(130.17, 82.89);

\path[draw=drawColor,line width= 0.2pt,line join=round] (135.58, 12.83) --
	(135.58, 82.89);

\path[draw=drawColor,line width= 0.2pt,line join=round] (140.98, 12.83) --
	(140.98, 82.89);

\path[draw=drawColor,line width= 0.2pt,line join=round] (146.39, 12.83) --
	(146.39, 82.89);

\path[draw=drawColor,line width= 0.2pt,line join=round] (151.80, 12.83) --
	(151.80, 82.89);

\path[draw=drawColor,line width= 0.2pt,line join=round] (157.21, 12.83) --
	(157.21, 82.89);

\path[draw=drawColor,line width= 0.2pt,line join=round] (162.61, 12.83) --
	(162.61, 82.89);
\definecolor{fillColor}{gray}{0.35}

\path[fill=fillColor] (  9.84, 16.02) rectangle ( 12.55, 55.36);

\path[fill=fillColor] ( 15.25, 16.02) rectangle ( 17.95, 44.99);

\path[fill=fillColor] ( 20.66, 16.02) rectangle ( 23.36, 58.93);

\path[fill=fillColor] ( 26.07, 16.02) rectangle ( 28.77, 66.53);

\path[fill=fillColor] ( 31.47, 16.02) rectangle ( 34.18, 38.28);

\path[fill=fillColor] ( 36.88, 16.02) rectangle ( 39.59, 58.10);

\path[fill=fillColor] ( 42.29, 16.02) rectangle ( 44.99, 55.79);

\path[fill=fillColor] ( 47.70, 16.02) rectangle ( 50.40, 44.70);

\path[fill=fillColor] ( 53.11, 16.02) rectangle ( 55.81, 63.10);

\path[fill=fillColor] ( 58.51, 16.02) rectangle ( 61.22, 72.03);

\path[fill=fillColor] ( 63.92, 16.02) rectangle ( 66.63, 76.34);

\path[fill=fillColor] ( 69.33, 16.02) rectangle ( 72.03, 59.87);

\path[fill=fillColor] ( 74.74, 16.02) rectangle ( 77.44, 60.61);

\path[fill=fillColor] ( 80.15, 16.02) rectangle ( 82.85, 70.20);

\path[fill=fillColor] ( 85.55, 16.02) rectangle ( 88.26, 34.46);

\path[fill=fillColor] ( 90.96, 16.02) rectangle ( 93.66, 53.86);

\path[fill=fillColor] ( 96.37, 16.02) rectangle ( 99.07, 58.76);

\path[fill=fillColor] (101.78, 16.02) rectangle (104.48, 53.99);

\path[fill=fillColor] (107.18, 16.02) rectangle (109.89, 58.23);

\path[fill=fillColor] (112.59, 16.02) rectangle (115.30, 16.02);

\path[fill=fillColor] (118.00, 16.02) rectangle (120.70, 45.87);

\path[fill=fillColor] (123.41, 16.02) rectangle (126.11, 66.52);

\path[fill=fillColor] (128.82, 16.02) rectangle (131.52, 47.92);

\path[fill=fillColor] (134.22, 16.02) rectangle (136.93, 40.33);

\path[fill=fillColor] (139.63, 16.02) rectangle (142.34, 16.02);

\path[fill=fillColor] (145.04, 16.02) rectangle (147.74, 57.65);

\path[fill=fillColor] (150.45, 16.02) rectangle (153.15, 16.02);

\path[fill=fillColor] (155.86, 16.02) rectangle (158.56, 54.02);

\path[fill=fillColor] (161.26, 16.02) rectangle (163.97, 40.95);
\end{scope}
\begin{scope}
\path[clip] (  0.00,  0.00) rectangle (165.86, 82.89);
\definecolor{drawColor}{gray}{0.30}

\node[text=drawColor,anchor=base east,inner sep=0pt, outer sep=0pt, scale=  0.60] at (  3.00, 14.06) {0};

\node[text=drawColor,anchor=base east,inner sep=0pt, outer sep=0pt, scale=  0.60] at (  3.00, 32.26) {2};

\node[text=drawColor,anchor=base east,inner sep=0pt, outer sep=0pt, scale=  0.60] at (  3.00, 50.46) {4};

\node[text=drawColor,anchor=base east,inner sep=0pt, outer sep=0pt, scale=  0.60] at (  3.00, 68.66) {6};
\end{scope}
\begin{scope}
\path[clip] (  0.00,  0.00) rectangle (165.86, 82.89);
\definecolor{drawColor}{gray}{0.30}

\node[text=drawColor,anchor=base,inner sep=0pt, outer sep=0pt, scale=  0.75] at ( 11.19,  3.00) {$g^3_1$};

\node[text=drawColor,anchor=base,inner sep=0pt, outer sep=0pt, scale=  0.75] at ( 22.01,  3.00) {$g^4_1$};

\node[text=drawColor,anchor=base,inner sep=0pt, outer sep=0pt, scale=  0.75] at ( 32.83,  3.00) {$g^4_3$};

\node[text=drawColor,anchor=base,inner sep=0pt, outer sep=0pt, scale=  0.75] at ( 43.64,  3.00) {$g^4_5$};

\node[text=drawColor,anchor=base,inner sep=0pt, outer sep=0pt, scale=  0.75] at ( 54.46,  3.00) {$g^5_1$};

\node[text=drawColor,anchor=base,inner sep=0pt, outer sep=0pt, scale=  0.75] at ( 65.27,  3.00) {$g^5_3$};

\node[text=drawColor,anchor=base,inner sep=0pt, outer sep=0pt, scale=  0.75] at ( 76.09,  3.00) {$g^5_5$};

\node[text=drawColor,anchor=base,inner sep=0pt, outer sep=0pt, scale=  0.75] at ( 86.90,  3.00) {$g^5_7$};

\node[text=drawColor,anchor=base,inner sep=0pt, outer sep=0pt, scale=  0.75] at ( 97.72,  3.00) {$g^5_9$};

\node[text=drawColor,anchor=base,inner sep=0pt, outer sep=0pt, scale=  0.75] at (108.54,  3.00) {$g^5_{11}$};

\node[text=drawColor,anchor=base,inner sep=0pt, outer sep=0pt, scale=  0.75] at (119.35,  3.00) {$g^5_{13}$};

\node[text=drawColor,anchor=base,inner sep=0pt, outer sep=0pt, scale=  0.75] at (130.17,  3.00) {$g^5_{15}$};

\node[text=drawColor,anchor=base,inner sep=0pt, outer sep=0pt, scale=  0.75] at (140.98,  3.00) {$g^5_{17}$};

\node[text=drawColor,anchor=base,inner sep=0pt, outer sep=0pt, scale=  0.75] at (151.80,  3.00) {$g^5_{19}$};

\node[text=drawColor,anchor=base,inner sep=0pt, outer sep=0pt, scale=  0.75] at (162.61,  3.00) {$g^5_{21}$};
\end{scope}
\end{tikzpicture}}\hfill
\subfloat[$G_{7208}\quad|V|=19,998\quad|E|=39,992$]{\label{fig:graphlet-frequency-d}\includegraphics[width=.16\textwidth,keepaspectratio]{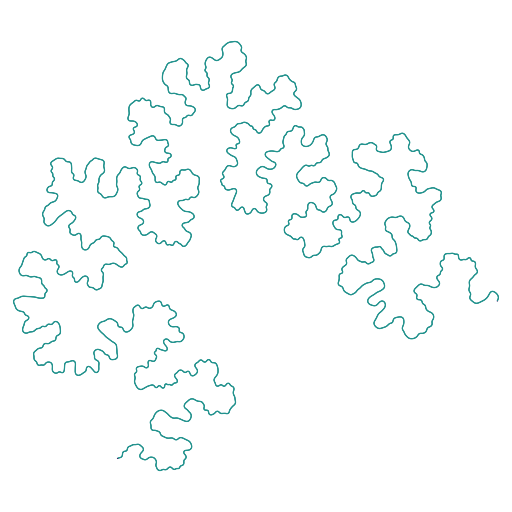}
\begin{tikzpicture}[x=1pt,y=1pt]
\definecolor{fillColor}{RGB}{255,255,255}
\path[use as bounding box,fill=fillColor,fill opacity=0.00] (0,0) rectangle (165.86, 82.89);
\begin{scope}
\path[clip] (  7.95, 12.83) rectangle (165.86, 82.89);
\definecolor{drawColor}{gray}{0.92}

\path[draw=drawColor,line width= 0.2pt,line join=round] (  7.95, 25.12) --
	(165.86, 25.12);

\path[draw=drawColor,line width= 0.2pt,line join=round] (  7.95, 43.31) --
	(165.86, 43.31);

\path[draw=drawColor,line width= 0.2pt,line join=round] (  7.95, 61.51) --
	(165.86, 61.51);

\path[draw=drawColor,line width= 0.2pt,line join=round] (  7.95, 79.71) --
	(165.86, 79.71);

\path[draw=drawColor,line width= 0.2pt,line join=round] (  7.95, 16.02) --
	(165.86, 16.02);

\path[draw=drawColor,line width= 0.2pt,line join=round] (  7.95, 34.21) --
	(165.86, 34.21);

\path[draw=drawColor,line width= 0.2pt,line join=round] (  7.95, 52.41) --
	(165.86, 52.41);

\path[draw=drawColor,line width= 0.2pt,line join=round] (  7.95, 70.61) --
	(165.86, 70.61);

\path[draw=drawColor,line width= 0.2pt,line join=round] ( 11.19, 12.83) --
	( 11.19, 82.89);

\path[draw=drawColor,line width= 0.2pt,line join=round] ( 16.60, 12.83) --
	( 16.60, 82.89);

\path[draw=drawColor,line width= 0.2pt,line join=round] ( 22.01, 12.83) --
	( 22.01, 82.89);

\path[draw=drawColor,line width= 0.2pt,line join=round] ( 27.42, 12.83) --
	( 27.42, 82.89);

\path[draw=drawColor,line width= 0.2pt,line join=round] ( 32.83, 12.83) --
	( 32.83, 82.89);

\path[draw=drawColor,line width= 0.2pt,line join=round] ( 38.23, 12.83) --
	( 38.23, 82.89);

\path[draw=drawColor,line width= 0.2pt,line join=round] ( 43.64, 12.83) --
	( 43.64, 82.89);

\path[draw=drawColor,line width= 0.2pt,line join=round] ( 49.05, 12.83) --
	( 49.05, 82.89);

\path[draw=drawColor,line width= 0.2pt,line join=round] ( 54.46, 12.83) --
	( 54.46, 82.89);

\path[draw=drawColor,line width= 0.2pt,line join=round] ( 59.87, 12.83) --
	( 59.87, 82.89);

\path[draw=drawColor,line width= 0.2pt,line join=round] ( 65.27, 12.83) --
	( 65.27, 82.89);

\path[draw=drawColor,line width= 0.2pt,line join=round] ( 70.68, 12.83) --
	( 70.68, 82.89);

\path[draw=drawColor,line width= 0.2pt,line join=round] ( 76.09, 12.83) --
	( 76.09, 82.89);

\path[draw=drawColor,line width= 0.2pt,line join=round] ( 81.50, 12.83) --
	( 81.50, 82.89);

\path[draw=drawColor,line width= 0.2pt,line join=round] ( 86.90, 12.83) --
	( 86.90, 82.89);

\path[draw=drawColor,line width= 0.2pt,line join=round] ( 92.31, 12.83) --
	( 92.31, 82.89);

\path[draw=drawColor,line width= 0.2pt,line join=round] ( 97.72, 12.83) --
	( 97.72, 82.89);

\path[draw=drawColor,line width= 0.2pt,line join=round] (103.13, 12.83) --
	(103.13, 82.89);

\path[draw=drawColor,line width= 0.2pt,line join=round] (108.54, 12.83) --
	(108.54, 82.89);

\path[draw=drawColor,line width= 0.2pt,line join=round] (113.94, 12.83) --
	(113.94, 82.89);

\path[draw=drawColor,line width= 0.2pt,line join=round] (119.35, 12.83) --
	(119.35, 82.89);

\path[draw=drawColor,line width= 0.2pt,line join=round] (124.76, 12.83) --
	(124.76, 82.89);

\path[draw=drawColor,line width= 0.2pt,line join=round] (130.17, 12.83) --
	(130.17, 82.89);

\path[draw=drawColor,line width= 0.2pt,line join=round] (135.58, 12.83) --
	(135.58, 82.89);

\path[draw=drawColor,line width= 0.2pt,line join=round] (140.98, 12.83) --
	(140.98, 82.89);

\path[draw=drawColor,line width= 0.2pt,line join=round] (146.39, 12.83) --
	(146.39, 82.89);

\path[draw=drawColor,line width= 0.2pt,line join=round] (151.80, 12.83) --
	(151.80, 82.89);

\path[draw=drawColor,line width= 0.2pt,line join=round] (157.21, 12.83) --
	(157.21, 82.89);

\path[draw=drawColor,line width= 0.2pt,line join=round] (162.61, 12.83) --
	(162.61, 82.89);
\definecolor{fillColor}{gray}{0.35}

\path[fill=fillColor] (  9.84, 16.02) rectangle ( 12.55, 56.75);

\path[fill=fillColor] ( 15.25, 16.02) rectangle ( 17.95, 16.02);

\path[fill=fillColor] ( 20.66, 16.02) rectangle ( 23.36, 58.26);

\path[fill=fillColor] ( 26.07, 16.02) rectangle ( 28.77, 56.08);

\path[fill=fillColor] ( 31.47, 16.02) rectangle ( 34.18, 54.19);

\path[fill=fillColor] ( 36.88, 16.02) rectangle ( 39.59, 16.02);

\path[fill=fillColor] ( 42.29, 16.02) rectangle ( 44.99, 16.02);

\path[fill=fillColor] ( 47.70, 16.02) rectangle ( 50.40, 16.02);

\path[fill=fillColor] ( 53.11, 16.02) rectangle ( 55.81, 59.87);

\path[fill=fillColor] ( 58.51, 16.02) rectangle ( 61.22, 57.05);

\path[fill=fillColor] ( 63.92, 16.02) rectangle ( 66.63, 48.87);

\path[fill=fillColor] ( 69.33, 16.02) rectangle ( 72.03, 16.02);

\path[fill=fillColor] ( 74.74, 16.02) rectangle ( 77.44, 16.02);

\path[fill=fillColor] ( 80.15, 16.02) rectangle ( 82.85, 16.02);

\path[fill=fillColor] ( 85.55, 16.02) rectangle ( 88.26, 16.02);

\path[fill=fillColor] ( 90.96, 16.02) rectangle ( 93.66, 57.13);

\path[fill=fillColor] ( 96.37, 16.02) rectangle ( 99.07, 16.02);

\path[fill=fillColor] (101.78, 16.02) rectangle (104.48, 16.02);

\path[fill=fillColor] (107.18, 16.02) rectangle (109.89, 16.02);

\path[fill=fillColor] (112.59, 16.02) rectangle (115.30, 51.66);

\path[fill=fillColor] (118.00, 16.02) rectangle (120.70, 16.02);

\path[fill=fillColor] (123.41, 16.02) rectangle (126.11, 16.02);

\path[fill=fillColor] (128.82, 16.02) rectangle (131.52, 16.02);

\path[fill=fillColor] (134.22, 16.02) rectangle (136.93, 16.02);

\path[fill=fillColor] (139.63, 16.02) rectangle (142.34, 16.02);

\path[fill=fillColor] (145.04, 16.02) rectangle (147.74, 16.02);

\path[fill=fillColor] (150.45, 16.02) rectangle (153.15, 16.02);

\path[fill=fillColor] (155.86, 16.02) rectangle (158.56, 16.02);

\path[fill=fillColor] (161.26, 16.02) rectangle (163.97, 16.02);
\end{scope}
\begin{scope}
\path[clip] (  0.00,  0.00) rectangle (165.86, 82.89);
\definecolor{drawColor}{gray}{0.30}

\node[text=drawColor,anchor=base east,inner sep=0pt, outer sep=0pt, scale=  0.60] at (  3.00, 14.06) {0};

\node[text=drawColor,anchor=base east,inner sep=0pt, outer sep=0pt, scale=  0.60] at (  3.00, 32.26) {2};

\node[text=drawColor,anchor=base east,inner sep=0pt, outer sep=0pt, scale=  0.60] at (  3.00, 50.46) {4};

\node[text=drawColor,anchor=base east,inner sep=0pt, outer sep=0pt, scale=  0.60] at (  3.00, 68.66) {6};
\end{scope}
\begin{scope}
\path[clip] (  0.00,  0.00) rectangle (165.86, 82.89);
\definecolor{drawColor}{gray}{0.30}

\node[text=drawColor,anchor=base,inner sep=0pt, outer sep=0pt, scale=  0.75] at ( 11.19,  3.00) {$g^3_1$};

\node[text=drawColor,anchor=base,inner sep=0pt, outer sep=0pt, scale=  0.75] at ( 22.01,  3.00) {$g^4_1$};

\node[text=drawColor,anchor=base,inner sep=0pt, outer sep=0pt, scale=  0.75] at ( 32.83,  3.00) {$g^4_3$};

\node[text=drawColor,anchor=base,inner sep=0pt, outer sep=0pt, scale=  0.75] at ( 43.64,  3.00) {$g^4_5$};

\node[text=drawColor,anchor=base,inner sep=0pt, outer sep=0pt, scale=  0.75] at ( 54.46,  3.00) {$g^5_1$};

\node[text=drawColor,anchor=base,inner sep=0pt, outer sep=0pt, scale=  0.75] at ( 65.27,  3.00) {$g^5_3$};

\node[text=drawColor,anchor=base,inner sep=0pt, outer sep=0pt, scale=  0.75] at ( 76.09,  3.00) {$g^5_5$};

\node[text=drawColor,anchor=base,inner sep=0pt, outer sep=0pt, scale=  0.75] at ( 86.90,  3.00) {$g^5_7$};

\node[text=drawColor,anchor=base,inner sep=0pt, outer sep=0pt, scale=  0.75] at ( 97.72,  3.00) {$g^5_9$};

\node[text=drawColor,anchor=base,inner sep=0pt, outer sep=0pt, scale=  0.75] at (108.54,  3.00) {$g^5_{11}$};

\node[text=drawColor,anchor=base,inner sep=0pt, outer sep=0pt, scale=  0.75] at (119.35,  3.00) {$g^5_{13}$};

\node[text=drawColor,anchor=base,inner sep=0pt, outer sep=0pt, scale=  0.75] at (130.17,  3.00) {$g^5_{15}$};

\node[text=drawColor,anchor=base,inner sep=0pt, outer sep=0pt, scale=  0.75] at (140.98,  3.00) {$g^5_{17}$};

\node[text=drawColor,anchor=base,inner sep=0pt, outer sep=0pt, scale=  0.75] at (151.80,  3.00) {$g^5_{19}$};

\node[text=drawColor,anchor=base,inner sep=0pt, outer sep=0pt, scale=  0.75] at (162.61,  3.00) {$g^5_{21}$};
\end{scope}
\end{tikzpicture}}
\caption{Examples of graphlet frequencies. 
The x-axis represents connected graphlets of size $k \in \{3, 4, 5\}$ and the y-axis represents the weighted frequency of each graphlet.
Four graphs are drawn with sfdp layouts \cite{Hu05}. 
If two graphs have similar graphlet frequencies, i.e., high topological similarity, they tend to have similar layout results (a and c). If not, the layout results look different (a and b).
However, in rare instances, two graphs can have similar graphlet frequencies (b and d), but vary in graph size, which might lead to different looking layouts.}
\label{fig:graphlet-frequency}
\vspace{-1em}
\end{figure*}

Graphlets are small, induced, and non-isomorphic subgraph patterns in a graph \cite{Przulj04} (\autoref{fig:graphlets}).
Graphlet frequencies (\autoref{fig:graphlet-frequency}) have been used to characterize biological networks \cite{Przulj04, Przulj07}, identify disease genes \cite{Milenkovic10}, and analyze social network structures \cite{Ugande13}.
Depending on the definition, the relative frequencies are called \emph{graphlet frequency distribution}, \emph{graphlet degree distribution}, or \emph{graphlet concentrations}.
While these works often have different definitions, the fundamental idea is to count the individual graphlets and compare their relative frequencies of occurrence between graphs.
A graph kernel based on graphlet frequencies, called a graphlet kernel, was first proposed by Shervashidze et al. \cite{Shervashidze09}.
The main idea is to use a graphlet frequency vector as the feature vector of a graph, then compute the similarity between graphs by defining the inner product of the feature vectors.

\section{Approach}
Our approach is a supervised learning of the relationship between topological features of existing graphs and their various layout results.
This also includes the layouts' aesthetic metrics.
Like many supervised learning methods, our approach requires empirical data (training data) of existing graphs, their layout results, and corresponding aesthetic metrics.
This generally takes a considerable amount of time, but can be considered a preprocessing step as it is only ever done once.
The benefit of machine learning is that as we add more graphs and their layout results, the performance generally improves.

\noindent
In this section, we introduce: 
\begin{enumerate}[topsep=0pt,itemsep=-1ex,partopsep=1ex,parsep=1ex]
\item A framework for designing better graphlet kernels
\item A process of using graph kernels to determine what a graph would look like in different layouts
\item A method to estimate the aesthetic metrics without calculating actual layouts
\end{enumerate}

\subsection{Graphlet Kernel Design Framework}
\label{sec:approach-framework}

One of the key challenges of our approach is choosing a graph kernel.
While many graph kernels are available, we focus on sampling based graphlet kernels because they are computationally efficient and are designed for unlabeled graphs.
To improve the performance of a graphlet kernel, we introduce a framework for designing graphlet kernels.
Our framework consists of three steps:
\begin{enumerate}[topsep=0pt,itemsep=-1ex,partopsep=1ex,parsep=1ex]
\item Sampling graphlet frequencies
\item Scaling graphlet frequency vectors
\item Defining an inner product between two graphlet frequency vectors
\end{enumerate}
For each step, we discuss several possible design choices.
Our framework defines several new types of graphlet kernels.

Existing studies related to graphlets and graphlet kernels are scattered throughout the literature, including machine learning and network science.
Each of the studies focuses on certain aspects of a graphlet kernel.
Our framework unifies the related works for graphlet kernels.

\subsubsection{Sampling Graphlet Frequencies}
\label{sec:approach-framework-sampling}
One of the challenges for constructing a graphlet kernel is computing the graphlet frequencies.
Exhaustive enumeration of graphlets with $k$ vertices in a graph $G$ is $O(|V|^k)$, which is prohibitively expensive, even for graphs with a few hundred or more vertices.
Thus, sampling approaches have been introduced to obtain the graphlet frequencies in a short amount of time with an acceptable error.

\textbf{Random vertex sampling (RV):}
Most existing works on graphlet kernels have sampled graphlets using a random vertex sampling method \cite{Borgwardt07}.
To sample a graphlet of size $k$, this method randomly chooses $k$ vertices in $V$ and induces the graphlet based on their adjacency in $G$.
This step is repeated until the number of sampled graphlets is sufficient.
Since this sampling method randomly chooses $k$ vertices without considering their interconnectivity in $G$, it has several limitations.
As many real-world networks are sparse \cite{Barabasi16}, $ |E| \ll O(|V|^2)$, most randomly sampled graphlets are disconnected.
Consequently, the frequency of disconnected graphlets would be much higher than connected ones.
If disconnected graphlets lack discriminating traits between graphs, and they outnumber the informative graphlets, comparing graphs becomes increasingly difficult.
While we could only sample connected graphlets by excluding disconnected ones, this requires a tremendous number of sampling iterations to sample a sufficient number of connected graphlets.
Since there is no lower bound on the number of iterations for sampling certain amounts of connected graphlets, using RV to sample only connected graphlets would lead to undesirable computation times.

\textbf{Random walk sampling (RW):}
There are other methods to sample graphlets based on random walks, such as Metropolis-Hasting random walk \cite{Rahman13}, subgraph random walk \cite{Wang14}, and expanded Markov chain \cite{Chen16}.
However, they have not been used for designing graphlet kernels in the machine learning community.
Unlike RV sampling, they sample graphlets by traversing the structure of a given graph.
That is, they search for the next sample vertices within the neighbors of the currently sampled vertices.
Thus, they are able to sample connected graphlets very well.

\subsubsection{Scaling Graphlet Frequency Vector}
\label{sec:approach-framework-scaling}
A graphlet frequency vector $\mathbf{x}$ is defined such that each component $x_i$ corresponds to the relative frequency of a graphlet $g_i$.
In essence, the graphlet frequency vector of a graph is the feature vector of the graph.

\textbf{Linear scale (\textsc{Lin}):} 
Many existing graphlet kernels use linear scaling, often called \emph{graphlet concentration}, which is the percentage of each graphlet in the graph.
As several works use weighted counts $w_i$ of each graphlet $g_i$, this scaling can be defined as:
\begin{equation*}
x_i = \frac{w_i}{\sum w_i}
\end{equation*}

\textbf{Logarithmic scale (\textsc{Log}):}
Similar to the vertex degree distribution, the distribution of graphlet frequency often exhibits a power-law distribution.
This again can cause a significant problem if graphlets that lack discriminating traits between graphs outnumber the informative graphlets.
Thus, several studies \cite{Przulj04, Rahman13} used a logarithmic scale of the graphlet frequency vector to solve this problem.
While the exact definitions of these methods differ, we generalize it using the following definition:
\begin{equation*}
x_i = \log \left(\frac{w_i + w_b}{\sum (w_i + w_b)} \right)
\end{equation*}
where $w_b > 0$ is a base weight to prevent $\log 0$.

\subsubsection{Defining Inner Product}
\label{sec:approach-framework-inner-product}
Several kernel functions can be used to define the inner product in a feature space $\pazocal{H}$.

\textbf{Cosine similarity (\textsc{Cos}):}
Most existing graphlet kernels use the dot product of two graphlet frequency vectors in Euclidean space, then normalize the kernel matrix.
This is equivalent to the cosine similarity of two vectors, which is the $L_2$-normalized dot product of two vectors:
\vspace{-0.45em}
\begin{equation*}
\left\langle \mathbf{x},\mathbf{x'} \right\rangle = \frac{\mathbf{x} \cdot \mathbf{x}^{\prime\mathsf{T}}}{\norm{\mathbf{x}}\norm{\mathbf{x'}}}
\end{equation*}

\textbf{Gaussian radial basis function kernel (\textsc{RBF}):}
This kernel is popularly used in various kernelized machine learning techniques: 
\vspace{-0.25em}
\begin{equation*}
\left\langle \mathbf{x},\mathbf{x'} \right\rangle = \exp\left(-\frac{\norm{\mathbf{x}-\mathbf{x'}}^{2}}{2\sigma^{2}}\right)
\vspace{-0.25em}
\end{equation*}
where $\sigma$ is a free parameter.

\textbf{Laplacian kernel (\textsc{Laplacian}):}
Laplacian kernel is a variant of RBF kernel: 
\vspace{-0.45em}
\begin{equation*}
\left\langle \mathbf{x},\mathbf{x'} \right\rangle = \exp\left(-\frac{\norm{\mathbf{x}-\mathbf{x'}}_1}{\sigma}\right)
\vspace{-0.05em}
\end{equation*}
where $\norm{x-x'}_1$ is the $L_1$ distance, or Manhattan distance, of the two vectors.

\subsection{What Would a Graph Look Like in This Layout? (WGL)}
\label{sec:approach-look}
Graph kernels provide us with the topological similarities between graphs.
Using these pairwise similarities, we design a nearest-neighbor based method, similar to $k$-nearest neighbors, to show what a graph would look like in different layouts.
Given a new input graph $G_{\text{input}}$, we find the $k$ most topologically similar graphs and show their existing layout results to the users.
Thus, if our assumption is true, users can expect the layout results of new input graphs by looking at the layout results of topologically similar graphs.

While graph kernels are able to find topologically similar graphs, many of them do not explicitly take the size of graphs into account.
In rare cases, it is possible to find topologically similar graphs that vary in size.
For instance, \autoref{fig:graphlet-frequency-b} and \autoref{fig:graphlet-frequency-d} have similar graphlet frequencies, yet have different layout results.
To prevent this, we add some constraints when we find similar graphs, such as only counting a graph that has more than half and less than double the number of vertices in the input graph.

As a result, for the new input graph $G_{\text{input}}$, we find $k$ most similar graphs as follows:
\begin{enumerate}[topsep=0pt,itemsep=-1ex,partopsep=1ex,parsep=1ex]
    \item Compute the similarity between existing graphs and $G_\text{input}$
    \item Remove the graphs that do not satisfy a set of constraints
    \item Select the $k$ most similar graphs
\end{enumerate}
After we have obtained the $k$ most similar graphs to the input graph $G_{\text{input}}$, we show their existing layout results to the user.

\subsection{Estimating Aesthetic Metrics (EAM)}
\label{sec:approach-metrics}
As the aesthetic metrics are continuous values, estimating the aesthetic metrics is a regression problem.
There are several kernelized regression models, such as Support Vector Regression (SVR) \cite{Smola04}, that can be used for estimating the aesthetic metrics based on the similarities between graphs obtained by graph kernels.
Computing actual aesthetic metrics of a layout requires a calculation of the layout first.
However, our approach is able to estimate the metrics without calculating actual layouts.

\textbf{Training:} 
To estimate the aesthetic metrics, we first need to train a regression model by:
\begin{enumerate}[topsep=0pt,itemsep=-1ex,partopsep=1ex,parsep=1ex]
    \item Prepare the training data (layouts and their aesthetic metrics of existing graphs)
    \item Compute a kernel matrix using a graph kernel (pairwise similarities between all graphs in the training data)
    \item Train regression model
\end{enumerate}

\textbf{Estimation:} To make an estimation of a new input graph, the following steps are required:
\begin{enumerate}[topsep=0pt,itemsep=-1ex,partopsep=1ex,parsep=1ex]
    \item Compute similarity between the input graph and other graphs in the training data 
    \item Estimate value using the trained regression model
\end{enumerate}

\section{Evaluation 1: Estimating Layout Aesthetic Metrics}
\label{sec:eval1}

There are several questions we wanted to answer within this evaluation:
\begin{itemize}[topsep=0pt,itemsep=-1ex,partopsep=1ex,parsep=1ex]
    \item Is our method able to accurately estimate the layout's aesthetic metrics without computing the layout?
    \item Is our method able to quickly obtain the estimations?
    \item Does our graph kernels, derived from our framework, outperform state-of-the-art graph kernels in terms of computation time and estimation accuracy?
\end{itemize}

\noindent
We describe the experimental design, apparatus, implementation, and metrics used in the study. We also answer each of these questions in the results section.

\subsection{Experimental Design }
We perform 10-fold cross-validations to compare 13 different graph kernels in terms of their accuracy and computation times for estimating four aesthetic metrics on eight layout methods.

\subsubsection{Datasets}
We started by collecting around 3,700 graphs from \cite{Davis11}, which includes, but not limited to, social networks, web document networks, and geometric meshes.
Without loss of generality, graphs with multiple connected components were broken down into separate graphs (one connected component to one graph).
After that, we removed any graph with less than 100 vertices, as there would be little benefit from a machine learning approach since most layout methods are fast enough for small graphs.
This left us with a total of 8,263 graphs for our study.
The graphs range from 100 vertices, and 100 edges up to 113 million vertices and 1.8 billion edges.
More details about the graphs, such as characteristic measures and layout results, can be found in \cite{Supplementary}.

Not all graphs were used for each layout method, as some layout algorithms failed to compute the results within a reasonable amount of time (10 days) or ran out of memory.
Exact numbers of graphs used for each layout are reported in \cite{Supplementary}.

\subsubsection{Kernels}
We compare a total of 13 graphlet kernels.
Using our framework, 12 graphlet kernels are derived from a combination of 2 graphlet sampling methods (RV and RW) $\times$ 2 types of graphlet frequency vector scaling (\textsc{Lin} and \textsc{Log}) $\times$ 3 inner products (\textsc{Cos}, \textsc{RBF}, and \textsc{Laplacian}).
We denote a kernel derived from our framework by a combination of the above abbreviations.
For example, \textsc{RW-Log-Laplacian} denotes a graphlet kernel which samples graphlets based on a random walk, uses a log scaled graphlet frequency vector, and computes the similarity using the Laplacian kernel function.
We also compare with state-of-the-art graphlet kernels.
The original graphlet kernel \cite{Shervashidze09} can be constructed using our framework and is included in the 12 kernels (\textsc{RV-Lin-Cos}).
Lastly, the $13^{\text{th}}$ graph kernel is a Deep Graphlet Kernel (DGK) \cite{Yanardag15}.

We used a sampling method from Chen et al.\cite{Chen16} as the RW sampling.
For all kernels, we sampled 10,000 graphlets of 3, 4, and 5 vertices for each graph. 
The graphlets of 3, 4, and 5 vertices are widely used due to computational costs. 
Also, the graphlets of 6 or more vertices are rare \cite{Rahman13}.
RW sampling considers only connected graphlets, as in \cite{Chen16},
while RV sampling counts both connected and disconnected graphlets, as in \cite{Shervashidze09, Yanardag15}.
All kernel matrices are normalized such that the similarity between a graph and itself has a value of 1.

\subsubsection{Layouts}
The process of laying out a graph has been actively studied for over five decades.
Several studies \cite{DiBattista94, DiBattista98, Landesberger11, Gibson12, Tamassia13} provide a comprehensive review of these layout methods.
While there are methods designed for specific purposes, such as orthogonal layout methods \cite{Kieffer16}, in this work, we focus on two-dimensional layout methods that draw all edges in straight lines.
Due to the volume of layout methods proposed, evaluating all the methods is impractical.
For this evaluation, we used eight representative layout methods of five families based on groupings found in \cite{Landesberger11, Gibson12}.
Our selection process prioritized methods that have been popularly used, have a publicly available implementation, and have a proven track record over other state-of-the-art methods.

\textbf{Force-directed methods:}
Force-directed methods are based on a physical model of attraction and repulsion.
They are among the first layout methods to be developed and are some of the most commonly used layout methods today.
In general, force-directed layout methods fall within two groups: spring-electrical \cite{Eades84, Fruchterman91, Frick94} and energy-based approaches \cite{Kamada89, Davidson96}.
We selected one from each group: Fruchterman-Reingold (FR) \cite{Fruchterman91} from spring-electrical approaches and Kamada-Kawai (KK) \cite{Kamada89} from energy-based approaches.

\textbf{Dimension reduction based method:}
Dimension reduction techniques, including multidimensional scaling (MDS) or principal component analysis (PCA), can be used to lay out a graph using the graph-theoretic distance between node pairs.
PivotMDS \cite{Brandes06} and High-Dimensional Embedder (HDE) \cite{Harel04} work by assigning several vertices as the pivots, then constructing a matrix representation with graph-theoretic distances of all vertices from the pivots. 
Afterward, dimension reduction techniques are applied to the matrix.
We selected HDE \cite{Harel04} in this family.

\textbf{Spectral method:}
Spectral layout methods use the eigenvectors of a matrix, such as the distance matrix \cite{Civril05} or the Laplacian matrix \cite{Koren05} of the graph, as coordinates of the vertices.
We selected the method by Koren \cite{Koren05} in this family.

\textbf{Multi-Level methods:}
Multilevel layout methods are developed to reduce computation time.
These multilevel methods hierarchically decompose the input graph into coarser graphs.
Then, they lay out the coarsest graph and use the vertex position as the initial layout for the next finer graph. 
This process is repeated until the original graph is laid out.
Several methods can be used to lay out the coarse graph, such as a force-directed method \cite{Walshaw03, Gajer02, Harel02, Hachul04, Hu05}, a dimension reduction based method \cite{Cohen97}, or a spectral method \cite{Koren02, Frishman07}.
We selected sfdp \cite{Hu05} and FM$^3$ \cite{Hachul04} in this family.

\textbf{Clustering based methods:}
Clustering based methods are designed to emphasize graph clusters in a layout.
When the graph size is large, users tend to ignore the number of edge crossings in favor of well-defined clusters \cite{VanHam08}.
We selected the Treemap based layout \cite{Muelder08B} and the Gosper curve based layout \cite{Muelder08} from this family, which utilizing the hierarchical clustering of a graph to lay out the graph.

\subsubsection{Aesthetic Metrics}
Aesthetic criteria, e.g., minimizing the number of edge crossings, are used for improving the readability of a graph layout.
Bennett et al. \cite{Bennett07} reviewed various aesthetic criteria from a perceptual basis.
Aesthetic metrics enable a quantitative comparison of the aesthetic quality of different layouts.
While there are many aesthetic criteria and metrics available, many of them are informally defined or are defined only for specific types of layout methods. 
Many also do not have a normalized metric for comparing graphs of different sizes, or are too expensive to compute (e.g., symmetry metric in \cite{Purchase02} is $O(n^7)$).
In this evaluation, we chose four aesthetic metrics because they have a normalized form, are not defined only for specific types of layouts, and can be computed in a reasonable amount of time.

\textbf{Crosslessness} ($m_c$) \cite{Purchase02}:
\textit{Minimizing the number of edge crossings} has been found as one of the most important aesthetic criteria in many studies \cite{Kieffer16, Purchase12, Huang07, Purchase02C}.
The crosslessness $m_c$ is defined as
\begin{equation*}
    m_c =
        \begin{cases}
            1 - \dfrac{c}{c_{\text{max}}},& \text{if}\ c_{\text{max}} > 0 \\
            1, & \text{otherwise}
        \end{cases}
\end{equation*}
where $c$ is the number of edge crossings and $c_{\text{max}}$ is the approximated upper bound of the number of edge crossings, which is defined as
\begin{equation*}
    \quad c_{\text{max}} = \frac{|E|\left(|E|-1\right)}{2} - \frac{1}{2}\sum\limits_{v \in V}\left(\text{deg}(v)\left(\text{deg}(v)-1\right)\right)\\
\end{equation*}

\textbf{Minimum angle metric} ($m_a$) \cite{Purchase02}:
This metric quantifies the criteria of \textit{maximizing the minimum angle between incident edges on a vertex}.
It is defined as the average absolute deviation of minimum angles between the incident edges on a vertex and the ideal minimum angle $\theta(v)$ of the vertex:
\begin{equation*}
    m_a =\ 1 - \frac{1}{|V|}\sum\limits_{v \in V} \left| \frac{\theta(v) - \theta_{\text{min}}(v)}{\theta(v)} \right|,\ \ \theta(v) =\ \frac{360^{\circ}}{\text{deg}(v)}
\end{equation*}
where $\theta_{\text{min}}(v_i)$ is the minimum angle between the incident edges on the vertex.

\textbf{Edge length variation} ($m_l$) \cite{Hachul07}:
\textit{Uniform edge lengths} have been found to be effective aesthetic criteria for measuring the quality of a layout in several studies \cite{Kieffer16}.
The coefficient of variance of the edge length ($l_{\text{cv}}$) has been used to quantify this criterion \cite{Hachul07}.
Since the upper bound of the coefficient of variation of $n$ values is $\sqrt{n - 1}$ \cite{Katsnelson57}, we divide $l_{\text{cv}}$ by $\sqrt{|E| - 1}$ to normalize:
\vspace{-1em}
\begin{equation*}
    m_l = \frac{l_{\text{cv}}}{\sqrt{|E|-1}},\ \ l_{\text{cv}} = \frac{l_{\sigma}}{l_{\mu}} =  \sqrt{\frac{\sum\limits_{e \in E} (l_e - l_{\mu})^2 }{|E| \cdot l_{\mu}^{\ 2}}}
\end{equation*}
where $l_{\sigma}$ is the standard deviation of the edge length and $l_{\mu}$ is the mean of the edge length.

\textbf{Shape-based metric} ($m_s$) \cite{Eades17}:
Shape-based metric is a more recent aesthetic metric and was proposed for evaluating the layouts of large graphs.
The shape-based metric $m_s$ is defined by the mean Jaccard similarity (MJS) between the input graph $G_{\text{input}}$ and the shape graph $G_{\text{S}}$:
\vspace{-1em}
\begin{equation*}
    m_s = \text{MJS}(G_{\text{input}}, G_{\text{S}}), \ \
    \text{MJS}(G_1, G_2) = \frac{1}{|V|} \sum\limits_{v \in V} \frac{|N_1(v) \cap N_2(v)|}{|N_1(v) \cup N_2(v)|}
\end{equation*}
where $G_1 = (V, E_1)$ and $G_2 = (V, E_2)$ are two graphs with the same vertex set and $N_i(u)$ is the set of neighbours of $v$ in $G_i$.
We use the Gabriel graph \cite{Gabriel69} as the shape graph.

\subsection{Apparatus and Implementation}
We perform 10-fold cross-validations of $\epsilon$-SVR implemented by \cite{LibSVM}.
To remove random effects of the fold assignments, we repeat the whole experiment 10 times and report mean accuracy metrics (\autoref{tab:eval1-accuracy}).

We obtained the implementation of DGK \cite{Yanardag15} from the authors.
The implementation of each layout method was gathered from:
sfdp \cite{graph-tool}, FM$^3$ \cite{OGDF}, FR \cite{graph-tool}, KK \cite{OGDF}, Spectral \cite{NetworkX}, and Treemap \cite{Muelder08B} and Gosper \cite{Muelder08} are provided by the authors.
Other kernels and layout methods were implemented by us.
For crosslessness ($m_c$), a GPU-based massively parallel implementation was used.
For other metrics, parallel CPU-based implementations written in C++ were used.
The machine we used to generate the training data and to conduct the experiment has two Intel Xeon processors (E5-4669 v4) with 22 cores (2.20 GHz) each, and two NVIDIA Titan X (Pascal) GPUs.

\subsection{Accuracy Metrics}
Root-Mean-Square Error (RMSE) measures the difference between measured values (ground truth) and estimated values by a model.
Given a set of measured values $\pazocal{Y} = \{y_1, \dots, y_n\}$ and a set of estimated values $\tilde{\pazocal{Y}} = \{\tilde{y_1}, \dots, \tilde{y_n}\}$, the RMSE is defined as:
\begin{equation*}
    \text{RMSE}(\pazocal{Y}, \tilde{\pazocal{Y}}) = \sqrt{\frac{1}{n} \sum\limits_{i}(y_i - \tilde{y}_i)^2}
\end{equation*}

The coefficient of determination ($R^2$) shows how well a model ``fits'' the given data.
The maximum $R^2$ score is 1.0 and it can have an arbitrary negative value.
Formally, it indicates the proportion of the variance in the dependent variable that is predictable from the independent variable, which is defined as:
\begin{equation*}
R^2(\pazocal{Y}, \tilde{\pazocal{Y}}) = 1 - \sum\limits_{i}(y_i - \tilde{y}_i)^2 \Big/ \sum\limits_{i}(y_i - y_{\mu})^2
\end{equation*}
where $y_{\mu}$ is the mean of measured values $y_i$.

\begin{table*}[ht]
\centering
\setlength{\tabcolsep}{0.5em} 
\caption{Estimation accuracy of the two most accurate kernels and the state-of-the-art kernels. 
We report Root-Mean-Square Error (RMSE) and the coefficient of determination ($R^2$) of estimation of four aesthetic metrics on eight layout methods.}
\label{tab:eval1-accuracy}
{\small
\begin{tabu}{|X[2,l] | X[.3,l] | X[1,c] | X[1,c] | X[1,c] | X[1,c] | X[1,c] | X[1,c] | X[1,c] | X[1,c] | X[1,c] | X[1,c] | X[1,c] | X[1,c] | X[1,c] | X[1,c] | X[1,c] | X[1,c]|}
\hline
\multicolumn2{|l|}{\multirow{2}{*}{Kernel}} & \multicolumn2{c|}{sfdp} & \multicolumn2{c|}{FM$^3$} & \multicolumn2{c|}{FR} & \multicolumn2{c|}{KK} & \multicolumn2{c|}{Spectral} & \multicolumn2{c|}{HDE} & \multicolumn2{c|}{Treemap} & \multicolumn2{c|}{Gosper} \\\cline{3-18}
\multicolumn2{|l|}{} & RMSE & $R^2$ & RMSE & $R^2$ & RMSE & $R^2$ & RMSE & $R^2$ & RMSE & $R^2$ & RMSE & $R^2$ & RMSE & $R^2$ & RMSE & $R^2$ \\\tabucline[1pt]{-}

\multirow{4}{*}{\shortstack[l]{Rank 1\\ \textsc{RW-Log-}\\ \textsc{Laplacian}}} & $m_c$ & $\mathbf{.0175}$ & $\mathbf{.9043}$ & $.0468$ & $.7319$ & $\mathbf{.0257}$ & $\mathbf{.8480}$ & $\mathbf{.0346}$ & $\mathbf{.8223}$ & $\mathbf{.1120}$ & $\mathbf{.6947}$ & $\mathbf{.0903}$ & $\mathbf{.8130}$ & $\mathbf{.0836}$ & $\mathbf{.6399}$ & $\mathbf{.0857}$ & $\mathbf{.6199}$\\\cline{2-18}
 & $m_a$ & $\mathbf{.1011}$ & $\mathbf{.8965}$ & $\mathbf{.1041}$ & $\mathbf{.8919}$ & $\mathbf{.0982}$ & $\mathbf{.9004}$ & $\mathbf{.1024}$ & $\mathbf{.8876}$ & $\mathbf{.1153}$ & $\mathbf{.8793}$ & $\mathbf{.1152}$ & $\mathbf{.8666}$ & $\mathbf{.1053}$ & $\mathbf{.8552}$ & $\mathbf{.1071}$ & $\mathbf{.8580}$\\\cline{2-18}
 & $m_l$ & $\mathbf{.0055}$ & $\mathbf{.9021}$ & $\mathbf{.0048}$ & $\mathbf{.8531}$ & $\mathbf{.0055}$ & $\mathbf{.9028}$ & $\mathbf{.0105}$ & $\mathbf{.4549}$ & $\mathbf{.0505}$ & $\mathbf{.6203}$ & $\mathbf{.0155}$ & $\mathbf{.5961}$ & $\mathbf{.0047}$ & $\mathbf{.8666}$ & $\mathbf{.0066}$ & $\mathbf{.8444}$\\\cline{2-18}
 & $m_s$ & $\mathbf{.0514}$ & $\mathbf{.9060}$ & $\mathbf{.0474}$ & $\mathbf{.9325}$ & $\mathbf{.0417}$ & $\mathbf{.8533}$ & $\mathbf{.0485}$ & $\mathbf{.9084}$ & $\mathbf{.0534}$ & $\mathbf{.9031}$ & $\mathbf{.0486}$ & $\mathbf{.8942}$ & $\mathbf{.0112}$ & $\mathbf{.8429}$ & $\mathbf{.0323}$ & $\mathbf{.7495}$\\\tabucline[1pt]{-}
\multirow{4}{*}{\shortstack[l]{Rank 2\\ \textsc{RW-Log-}\\ \textsc{RBF}}} & $m_c$ & $.0176$ & $.9036$ & $\mathbf{.0446}$ & $\mathbf{.7568}$ & $.0279$ & $.8218$ & $.0350$ & $.8182$ & $.1138$ & $.6845$ & $.0917$ & $.8072$ & $.0841$ & $.6356$ & $.0882$ & $.5976$\\\cline{2-18}
 & $m_a$ & $.1070$ & $.8840$ & $.1102$ & $.8788$ & $.1023$ & $.8920$ & $.1061$ & $.8793$ & $.1193$ & $.8706$ & $.1202$ & $.8546$ & $.1101$ & $.8416$ & $.1125$ & $.8434$\\\cline{2-18}
 & $m_l$ & $.0062$ & $.8793$ & $.0050$ & $.8412$ & $.0059$ & $.8874$ & $.0106$ & $.4497$ & $.0519$ & $.5992$ & $.0167$ & $.5291$ & $.0052$ & $.8417$ & $.0073$ & $.8127$\\\cline{2-18}
 & $m_s$ & $.0556$ & $.8900$ & $.0542$ & $.9116$ & $.0459$ & $.8227$ & $.0547$ & $.8833$ & $.0576$ & $.8875$ & $.0537$ & $.8708$ & $.0116$ & $.8299$ & $.0323$ & $.7491$\\\tabucline[1pt]{-}
\multirow{4}{*}{\shortstack[l]{Rank 11\\ \textsc{RV-Lin-}\\ \textsc{Cos} \cite{Shervashidze09}}} & $m_c$ & $.0387$ & $.5312$ & $.0771$ & $.2716$ & $.0577$ & $.2364$ & $.0783$ & $.0916$ & $.1533$ & $.4280$ & $.1770$ & $.2827$ & $.1324$ & $.0978$ & $.1336$ & $.0763$\\\cline{2-18}
 & $m_a$ & $.2883$ & $.1581$ & $.2907$ & $.1570$ & $.2817$ & $.1805$ & $.2850$ & $.1292$ & $.3019$ & $.1723$ & $.2978$ & $.1080$ & $.2688$ & $.0557$ & $.2726$ & $.0801$\\\cline{2-18}
 & $m_l$ & $.0168$ & $.0972$ & $.0121$ & $.0561$ & $.0169$ & $.0895$ & $.0138$ & $.0609$ & $.0812$ & $.0200$ & $.0239$ & $.0403$ & $.0116$ & $.2026$ & $.0156$ & $.1378$\\\cline{2-18}
 & $m_s$ & $.1721$ & $-.0552$ & $.1904$ & $-.0890$ & $.0984$ & $.1850$ & $.1653$ & $-.0656$ & $.1777$ & $-.0729$ & $.1538$ & $-.0606$ & $.0246$ & $.2373$ & $.0628$ & $.0521$\\\tabucline[1pt]{-}
\multirow{4}{*}{\shortstack[l]{Rank 12\\ DGK \cite{Yanardag15}}} & $m_c$ & $.0399$ & $.5029$ & $.0783$ & $.2500$ & $.0583$ & $.2207$ & $.0803$ & $.0448$ & $.1564$ & $.4041$ & $.1804$ & $.2541$ & $.1358$ & $.0489$ & $.1345$ & $.0630$\\\cline{2-18}
 & $m_a$ & $.2891$ & $.1536$ & $.2924$ & $.1467$ & $.2837$ & $.1690$ & $.2862$ & $.1217$ & $.3052$ & $.1537$ & $.3003$ & $.0930$ & $.2716$ & $.0357$ & $.2754$ & $.0612$\\\cline{2-18}
 & $m_l$ & $.0175$ & $.0246$ & $.0126$ & $-.0134$ & $.0177$ & $.0047$ & $.0140$ & $.0294$ & $.0811$ & $.0203$ & $.0243$ & $.0018$ & $.0128$ & $.0029$ & $.0185$ & $-.3883$\\\cline{2-18}
 & $m_s$ & $.1756$ & $-.0982$ & $.1928$ & $-.1171$ & $.1077$ & $.0236$ & $.1676$ & $-.0953$ & $.1807$ & $-.1094$ & $.1550$ & $-.0771$ & $.0286$ & $-.0846$ & $.0682$ & $-.1187$\\\hline

\end{tabu}
}

\end{table*}

\begin{figure*}[htb]
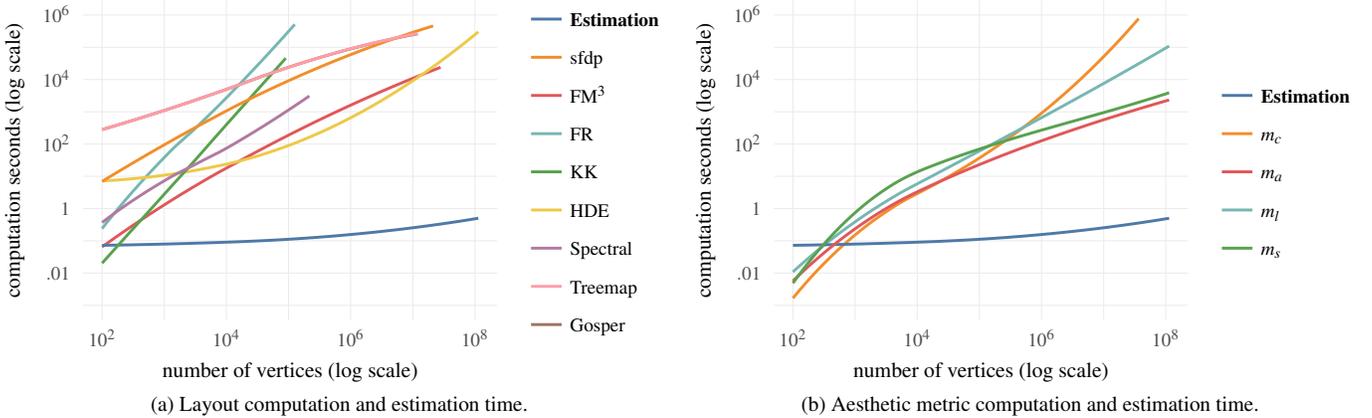

\centering
\vspace{-1.5em}
\subfloat[Layout computation and estimation time.]{\input{figures/eval1_perf_layout.tex}}\hfill
\subfloat[Aesthetic metric computation and estimation time.]{\input{figures/eval1_perf_metric.tex}}
\vspace{-0.5em}
\caption{Computation time results in log scale.
The plots show estimation times for our \textsc{RW-Log-Laplican} kernel, which has the highest accuracy. 
The plot on the left shows layout computation time while the plot on the right shows aesthetic metric computation time. 
As the number of vertices increases, the gap between our estimation and layout methods enlarges in both layout time and aesthetic metric computation time. 
Some layout methods could not be run on all graphs due to computation time and memory limitation of the implementation.
Treemap and Gosper overlap in the layout plot because the majority of the computation time is spent on hierarchical clustering.}
\vspace{-0.5em}
\label{fig:eval1-time}
\end{figure*}

\subsection{Results}
We report the accuracy and computation time for estimating the aesthetic metrics.

\subsubsection{Estimation Accuracy}
Due to space constraints, we only reported the results of the two most accurate kernels and state-of-the-art kernels \cite{Shervashidze09, Yanardag15} in \autoref{tab:eval1-accuracy}.
The results shown are mean RMSE (lower is better) and mean $R^2$ (higher is better) from 10 trials of 10-fold cross-validations.
The standard deviations of RMSE and $R^2$ are not shown because the values are negligible: all standard deviations of RMSE are lower than .0006, and all standard deviations of $R^2$ are lower than .0075.
We ranked the kernels based on the mean RMSE of all estimations.

The most accurate kernel is our \textsc{RW-Log-Laplacian} kernel.
Except for the crosslessness ($m_c$) of FM$^3$, our \textsc{RW-Log-Laplacian} kernel shows best estimation results in both RMSE and $R^2$ score for all four aesthetic metrics on all eight layout methods.
The second most accurate kernel is our \textsc{RW-Log-RBF} kernel and is the best for crosslessness ($m_c$) of FM$^3$.

Our \textsc{RW-Log-Laplacian} kernel (mean RMSE $=.0557$ and mean $R^2=.8169$) shows an average of 2.46 times lower RMSE than existing kernels we tested.
The original graphlet kernel \cite{Shervashidze09} (mean RMSE $=.1366$ and mean $R^2=.1216$), which is denoted as \textsc{RV-Lin-Cos}, ranked $11^{\text{th}}$.
The DGK \cite{Yanardag15} (mean RMSE $=.1388$ and mean $R^2=.0540$) ranked $12^{\text{th}}$.

Within the kernels we derived, the kernels using RW sampling show higher accuracy (mean RMSE $=.0836$ and mean $R^2 = .6247$) than ones using RV sampling (mean RMSE $=.1279$ and mean $R^2 = .2501$). 
The kernels using \textsc{Log} scaling show higher accuracy (mean RMSE $=.0955$ and mean $R^2 = .5279$) than ones using \textsc{Lin} (mean RMSE $=.1160$ and mean $R^2 = .3469$).
The kernels using \textsc{Laplacian} as the inner product show higher accuracy (mean RMSE $=.0956$ and mean $R^2 = .5367$) than ones using \textsc{RBF} (mean RMSE $=.1047$ and mean $R^2 = .4393$) and \textsc{Cos} (mean RMSE $=.1169$ and mean $R^2 = .3362$).

\subsubsection{Computation Time}
Since some algorithms are implemented in parallel while others are not, we report CPU times.
The estimation time is comprised of the computation steps required for estimation in \autoref{sec:approach-metrics}.
Our \textsc{RW-Log-Laplacian} kernel, which shows the best estimation accuracy, also shows the fastest computation time for estimation. 
On average, it takes .14093 seconds ($SD=1.9559$) per graph to make the estimations.
\autoref{fig:eval1-time} shows the computation time for layouts and their aesthetic metrics.

Most of the time spent on estimation was devoted to sampling graphlets.
The RW sampling \cite{Chen16} shows fastest computation time, with an average of .14089 seconds ($SD=1.9559$) per graph.
The graphlet sampling for DGK take longer than RW, with an average of 3.38 seconds ($SD=7.88$) per graph.
The RV sampling take the longest time, with an average of 6.81 seconds ($SD=7.04$).

\subsection{Discussion}
Our graph kernels perform exceptionally well in both estimation accuracy and computation time. 
Specifically, \textsc{RW-Log-Laplacian} outperforms all other kernels in all metrics except crosslessness ($m_c$) on the FM$^3$. 
\textsc{RW-Log-RBF} is the best performing kernel on FM$^3$'s crosslessness ($m_c$) and is the second best kernel.
Existing graph kernels are ranked in the bottom three of all kernel methods we tested.

A possible explanation for this is that certain types of graphlets are essential for accurate estimation.
The kernels using RW sampling, which samples connected graphlets very efficiently, show a higher accuracy than other kernels.
While disconnected graphlets are shown to be essential for classification problems in bioinformatics \cite{Shervashidze09}, for our problem, we suspect that connected graphlets are more important for accurate estimation.

Other sampling methods are not suitable for sampling connected graphlets.
There are 49 possible graphlets, where the size of each graphlet is $k \in \{3, 4, 5\}$, and 29 of them are connected graphlets (\autoref{fig:graphlets}).
However, when we sample 10,000 graphlets per graph using RV sampling, on average only 1.913\% ($SD=6.271$) of sampled graphlets are connected graphlets.
Furthermore, 35.77\% of graphs have no connected graphlets in the samples even though all graphs are connected, making it impossible to effectively compare graphs.

The kernels using \textsc{Log} scaling show more accurate estimations than the ones using \textsc{Lin}.
We suspect this is because of the distribution of graphlets, which often exhibit a power-law distribution.
Thus, when using \textsc{Log} scaling, a regression model is less affected by the graphlets with overwhelming frequencies and becomes better at discriminating graphs with different structures.

Our estimation times are fast and scale well, as shown in \autoref{fig:eval1-time}. 
At around 200 vertices, our estimation times become faster than all other layout computation times.
Our estimation times also outperform the computations of the four aesthetic metrics past 1,000 vertices.
As the size of a graph increases, the differences become larger, to the point that our \textsc{RW-Log-Laplacian} takes several orders of magnitude less time than both layout computation and metric computation.
Normally the layout has to be calculated in order to calculate aesthetic metrics.
This is not the case for our estimation as the aesthetic metrics can be estimated without the layout result, leading to a considerable speed up.

It is interesting to see that each sampling method shows different computation times, even though we sampled the same amount of graphlets.
A possible explanation for this discrepancy can be attributed to locality of reference.
We stored the graphs using an adjacency list data structure in memory.
Thus, RW sampling, which finds the next sample vertices within the neighbors of currently sampled vertices, tends to exhibit good locality.
On the other hand, RV sampling chooses vertices randomly, thus it would show poor locality which leads to cache misses and worse performance.
The sampling method of DGK is a variant of RV. 
After one graphlet is randomly sampled, its immediate neighbors are also sampled.
This would have better locality than RV and could explain the better computation time.

In terms of training, except for DGK, all other kernels spend negligible times computing the kernel matrix, with an average of 5.99 seconds ($SD=3.26$).
However, computing the kernel matrix of DGK takes a considerable amount of time because it requires computation of language modeling (implemented by \cite{Gensim}).
On average it take 182.96 seconds ($SD=9.31$).

Since there are many parameters of each layout method, and most parameters are not discrete, it is impossible to test all combinations of parameter settings.
To simplify the study, we only used default parameters for each layout method.
It is possible to apply our approach to different predefined parameter settings on the same layout method.
However, generating new training data for multiple settings can be time consuming.

\begin{figure}[ht]
\centering
\includegraphics[width=250pt, keepaspectratio]{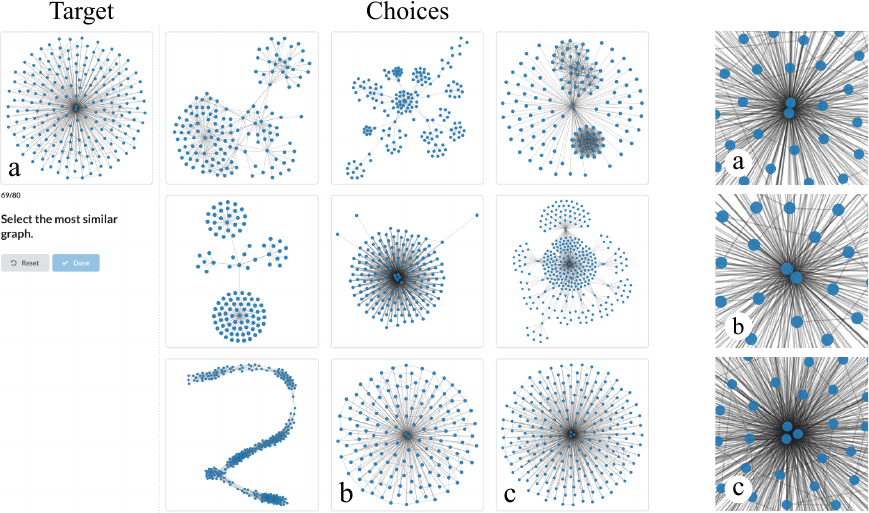}
\vspace{-1em}
\caption{A task from the user study. 
For each task, the participants were given one target graph and nine choice graphs. 
They were then asked to rank the three most similar graphs in order of decreasing similarity.
The three images on the right show the central nodes of (a), (b), and (c).}
\vspace{-1em}
\label{fig:task}
\end{figure}

\begin{figure*}[htb]
\centering
\input{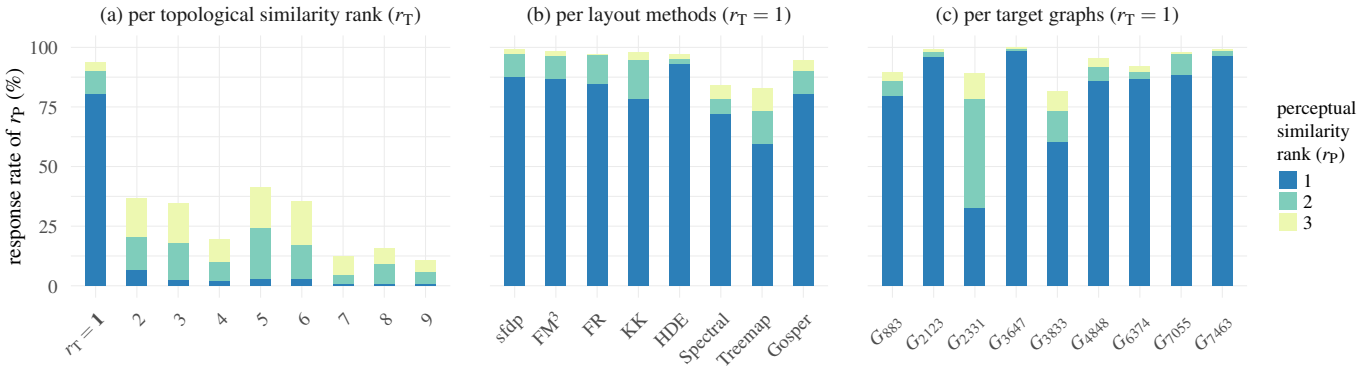}
\vspace{-2.25em}
\caption{Summary of the user study results. 
(a) response rate of perceptual similarity rank ($r_{\text{P}}$) for each topological similarity rank ($r_{\text{T}}$).
The plot in the middle (b) shows the response rate on topological similarity rank of 1 per layout method while the plot on the right (c) shows per target graph.
}
\label{fig:eval2-response-rate}
\end{figure*}

\setlength{\tabcolsep}{0.22em} 
\begin{table*}[htb]
\caption{Descriptive statistics of perceptual similarity rank ($r_{\text{P}}$). $\mu$: mean, $SD$: standard deviation, $\tilde{r_{\text{P}}}$: median, and $IQR$: interquartile range.
Our predicted choices are ranked on average 1.35 by the participants.}
\label{tab:eval2-desc-stats}
\vspace{-1em}
\centering
{\small
\subfloat[per topological similarity rank ($r_{\text{T}}$)]{%
\label{tab:eval2-desc-stats:topo}
\begin{tabu} spread 0pt
{|X[1,r]|X[1,c]|X[1,c]|X[1,c]|X[1,c]|X[1,c]|X[1,c]|X[1,c]|X[1,c]|X[1,c]|} \hline
        & \textbf{1}    & 2    & 3    & 4    & 5    & 6    & 7    & 8    & 9 \\\hline 
                 $\mu$ & \textbf{1.35} & 3.36 & 3.45 & 3.68 & 3.31 & 3.45 & 3.82 & 3.74 & 3.83 \\
                  $SD$ & .82           & .96  & .84  & .71   & .91   & .84  & .53  & .63  & .53  \\
$\tilde{r_{\text{P}}}$ & \textbf{1}    & 4    & 4    & 4    & 4    & 4    & 4    & 4    & 4    \\
                 $IQR$ & 0             & 1    & 1    & 0    & 1    & 1    & 0    & 0    & 0    \\
\hline 
\end{tabu}}\hfill
\subfloat[per layout methods ($r_{\text{T}} = 1$)]{%
\label{tab:eval2-desc-stats:layout}
\begin{tabu} spread 0pt {|X[1,c]|X[1,c]|X[1,c]|X[1,c]|X[1,c]|X[1,c]|X[1,c]|X[1,c]|} \hline
sfdp & FM$^3$ & FR   & KK   & HDE  & Spct. & Tree. & Gos. \\\hline 
1.16 & 1.19   & 1.21 & 1.29 & 1.14 & 1.65  & 1.84 & 1.35 \\
.46  & .55    & .6   & .63  & .58  & 1.14  & 1.17  & .81   \\
1    & 1      & 1    & 1    & 1    & 1     & 1    & 1    \\
0    & 0      & 0    & 0    & 0    & 1     & 2    & 0    \\
\hline 
\end{tabu}}\hfill
\subfloat[per target graphs ($r_{\text{T}} = 1$)]{%
\label{tab:eval2-desc-stats:target}
\begin{tabu} spread 0pt {|X[1,c]|X[1,c]|X[1,c]|X[1,c]|X[1,c]|X[1,c]|X[1,c]|X[1,c]|X[1,c]|} \hline
$G_{883}$ & $G_{2123}$ & $G_{2331}$ & $G_{3647}$ & $G_{3833}$ & $G_{4848}$ & $G_{6374}$ & $G_{7055}$ & $G_{7463}$ \\\hline 
1.45  & 1.07  & 2   & 1.02  & 1.85  & 1.27  & 1.32  & 1.17  & 1.06 \\
.98   & .36   & .94 & .17   & 1.18  & .74    & .86   & .53   & .34  \\
1     & 1     & 2   & 1     & 1     & 1     & 1     & 1     & 1    \\
0     & 0     & 1   & 0     & 2     & 0     & 0     & 0     & 0    \\
\hline 
\end{tabu}}
}
\vspace{-1em}
\end{table*}

\section{Evaluation 2: What Would a Graph Look Like in This Layout?}
\label{sec:eval2}

In this section, we describe our user study that evaluates how well our WGL method (\autoref{sec:approach-look}) is able to find graphs that users assess as being perceptually similar to the actual layout results.

\subsection{Experimental Design}
We designed a ranking experiment to compare topological similarity ranks ($r_{\text{T}}$) obtained by our WGL method and perceptual similarity ranks ($r_{\text{P}}$) assessed by humans.
That is, if both our WGL method and participants' choices match, then we can conclude our WGL method is able to find perceptually similar graphs to the actual layout results.

\subsubsection{Task}
For each task, participants were given one target graph and nine choice graphs.
An example of a task given to participants is shown in \autoref{fig:task}. 
They were asked to rank the three most similar graphs to the target graph in order of decreasing similarity.
The instructions were given as such, ``select the most similar graph.'' for the first choice and ``select the next most similar graph.'' for the second and third choice.
To avoid biases, we did not specify what is ``similar'' and let the participants decide for themselves.
In each task, the same layout method was used for target and choice graphs. 
We did not notify this to the participants.
To test our approach using graphs with different topological structures and different layout methods, 
we used 72 tasks comprised of nine graphs and eight layout methods.

While other task designs are possible, such as participants ranking all nine graphs or individually rating all nine choices, during our pilot study we found that these task designs are overwhelming to the participants.
The participants found it very difficult to rank between dissimilar graphs. 
For example, in \autoref{fig:task}, selecting the fourth or fifth most similar graph is challenging as there is little similarity to the target graph after the first three choices.
Also, the focus of our evaluation is to determine how well our WGL method is able to find similar graphs, not dissimilar graphs.
Thus, we only need to see which graphs our participants perceive as similar in the task, which simultaneously reduces task load when compared to ranking all nine graphs.

\subsubsection{Graphs}
To gather an unbiased evaluation, the target graphs and their choice graphs must be carefully selected.

\textbf{Selecting target graphs.}
To test our approach on graphs with different topological structures, we select nine target graphs as follows:
We clustered the 8,263 graphs into nine groups using spectral clustering \cite{SpectralClustering}.
For each cluster, we select ten representative graphs which have the highest sum of topological similarity within the cluster (i.e., the ten nearest graphs to the center of each cluster). 
Then, we randomly select one graph from these ten representative graphs to be the target graph.
\autoref{fig:teaser} shows the selected nine target graphs in FM$^3$ layout \cite{Hachul04}.

\textbf{Selecting choice graphs.}
To test our approach on graphs with different topological similarities obtained by graph kernels, we select nine choice graphs for each target graph as follows:
We compute nine clusters of graphs based on the topological similarity between a graph and the target graph using Jenks natural breaks \cite{Jenks67}, which can be seen as one dimensional $k$-means clustering.
We designate one of the nine choice graphs to represent what our approach would predict. This is done by selecting the graph with the highest similarity to the target graph from the cluster nearest to the target graph.
For the other eight clusters, we randomly select one choice graph from a set of ten graphs that have the highest sum of similarity within the cluster (i.e., the ten nearest graphs to the center of each cluster).
These can be considered as representative graphs of each cluster.
A list of the selected nine choice graphs for each target graph can be found in \cite{Supplementary}.

When we select a choice graph, we filter out graphs that have less than half or more than double the amount of vertices as the target graph.
This can be considered as a constraint in our WGL method \autoref{sec:approach-look}.
To evaluate under the condition that isomorphic graphs to the target graph are not present in the training data, we also filter out graphs that have the same number of vertices and edges as the target graph. 

The clustering approach based on topological similarity can be used for defining a topological classification of graphs. 
However, it would require further analysis of resultant clusters.

\subsection{Apparatus}
We used our \textsc{RW-Log-Laplacian} kernel to compute the topological similarity between graphs.
Although other kernels can be used, we decided on one that shows the highest estimated accuracy in the first evaluation and provides the best chance to see if a kernel can find perceptually similar graphs.

The experiment was conducted on an iPad Pro which has a 12.9 inch display with 2,732$\times$2,048 pixels. 
Each of the graphs were presented in 640$\times$640 pixels.
All vertices  were drawn using the same blue color (Red: .122, Green: .467, Blue: .706, Alpha: .9) and edges were drawn using dark grey color (R: .1, G: .1, B: .1, A: .25), as shown in \autoref{fig:task}. 

\subsection{Procedure}
Prior to the beginning of the experiment, the participants were asked several questions about demographic information, such as age and experience with graph visualization.
To familiarize participants with the system, eight training tasks were given, at which time they were allowed to ask any questions to the moderator. 
Once the training was done, the moderator did not communicate with the participant.
The 72 tasks were presented in randomized order to the participants. 
The nine choice graphs of each task were also presented in a randomized order.
For each task, we asked the participants to briefly explain why he or she selected the particular graph (think aloud protocol).

\subsection{Participants}
We recruited 30 (9 females and 21 males) participants for our user study.
The ages of the participants ranged from 18 to 36 years, with the mean age of 26.67 years (SD = 3.68).
Most of the participants were from science and engineering backgrounds: 22 computer science, 2 electrical and computer engineering, 2 cognitive science, 1 animal science, 1 political science, and 1 literature.
28 participants indicated that they had seen a graph visualization (e.g., a node-link diagram), 21 had used one, and 16 had created one before.
On average, each participant took 28.94 minutes ($SD = 11.49$) to complete all 72 tasks.

\subsection{Results}
For each task, the nine choices receive a topological similarity rank ($r_{\text{T}}$) from one to nine in order of decreasing topological similarity.
We define predicted choice as the choice graph that our method ranks as the most topologically similar to the target graph ($r_{\text{T}}=1$).
Based on the responses by the participants, the three choices receive a perceptual similarity rank ($r_{\text{P}}$) from one to three, where one is the first chosen graph.
Choices not selected by the participants are ranked $r_{\text{P}}=4$.
Results of our evaluation can be found in \autoref{fig:eval2-response-rate}.

Overall, 80.46\% of the time, participants' responses for rank one ($r_{\text{P}}=1$) match our predicted choices ($r_{\text{T}}=1$).
90.27\% of participants' responses within rank one and rank two ($r_{\text{P}} = \{1, 2\}$) contain the predicted choices, and 93.80\% of responses within ranks one, two, and three ($r_{\text{P}} = \{1, 2, 3\}$) contain the predicted choice.
A Friedman test (non-parametric alternative to one-way repeated measures ANOVA) shows a significant effect of the topological similarity rankings ($r_{\text{T}}$) on the perceptual similarity rankings ($r_{\text{P}}$) with $\chi^2(8)=6343.9,\ p < .0001$.
The mean $r_{\text{P}}$ of the predicted choices is 1.35 ($SD=.82$ and $IQR=0$), which is clearly different than other choices ($r_{\text{T}} > 1$) as shown in \autoref{tab:eval2-desc-stats}.
Post-hoc analysis with Wilcoxon signed-rank tests using Bonferroni correction confirms that the predicted choices are ranked significantly higher ($p < .0001$) than all other choices.

To see the effect layout methods have on participants' responses, we break down the responses with a topological similarity rank of one ($r_{\text{T}} = 1$), or predicted choice, by each layout method.
Except for Spectral and Treemap, the predicted choices are ranked in more than 78.52\% (up to 93.33\%) of participants' responses as being the most perceptually similar. 
In more than 94.44\% (up to 99.26\%) of participants' responses, the predicted choices are within the three most perceptually similar graphs ($r_{\text{P}} = \{1, 2, 3\}$), as shown in \autoref{fig:eval2-response-rate}b.
For Spectral, the predicted choices are ranked in 72.22\% of participants' responses as being the most similar graph, and 84.07\% of responses as being within the three most similar graphs.
For Treemap, the predicted choices are ranked in 59.63\% of participants' responses as the most similar graph, and 82.59\% of responses as being within the three most similar graphs.
A Friedman test shows a significant effect of layout method on perceptual similarity rankings ($r_{\text{P}}$) with $\chi^2(7)=200.85,\ p < .0001$. 
Except for Spectral and Treemap, the mean $r_{\text{P}}$ of the predicted choices for each layout method is close to 1, from 1.16 to 1.35 ($SD=.46$--$.81$ and $IQR=0$), as shown in \autoref{tab:eval2-desc-stats}b.
The means $r_{\text{P}}$ of Spectral and Treemap are 1.65 ($SD=1.14$ and $IQR=1$) and 1.84 ($SD=1.17$ and $IQR=2$), respectively. 
Post-hoc analysis shows that the predicted choices with Treemap are ranked significantly lower than the predicted choices with other layout methods ($p < .0001$) except for Spectral. 
The predicted choices with Spectral are also ranked by participants as being significantly lower than the predicted choices with other layout methods ($p < .05$) except for Gosper and Treemap.

We also break down the responses for the topological similarity rank of one ($r_{\text{T}} = 1$) by each target graph.
Except for $G_{2331}$ and $G_{3833}$, the predicted choices are ranked in more than 79.58\% (up to 98.33\%) of responses as being the most similar, and more than 92.08\% (up to 99.99\%) of responses as being within the three most similar graphs ($r_{\text{P}} = \{1, 2, 3\}$), as shown in \autoref{fig:eval2-response-rate}c.
For $G_{2331}$, the predicted choices are ranked in 32.92\% of all responses as being the most similar graph, 78.33\% of responses as being within the two most similar graphs, and 89.16\% of responses as being within the three most similar graphs.
For $G_{3833}$, the predicted choices are ranked in 60.42\% of participants' responses as being the most similar graph, 73.33\% of responses as being within the two most similar graphs, and 81.67\% of responses as being within the three most similar graphs.
A Friedman test shows a significant effect of target graphs on perceptual similarity rankings ($r_{\text{P}}$) with $\chi^2(8)=511,\ p < .0001$. 
Except for $G_{2331}$ and $G_{3833}$, the mean $r_{\text{P}}$ of the predicted choices for each target graph is close to 1, from 1.06 to 1.45 ($SD=.34$--$.98$ and $IQR=0$), as shown in \autoref{tab:eval2-desc-stats}b.
The means $r_{\text{P}}$ of $G_{2331}$ and $G_{3833}$ are 2 ($SD=.94$ and $IQR=1$) and 1.85 ($SD=1.18$ and $IQR=2$), respectively. 
Post-hoc analysis shows that the predicted choices of $G_{2331}$ and $G_{3833}$ are ranked significantly lower than predicted choices of other target graphs ($p < .0001$).

\subsection{Discussion}
The results of the user study show that in more than 80\% of participants' responses, the predicted choices are ranked as being the most perceptually similar graph to the target graphs.
Also, more than 93\% of the responses ranked the predicted choices as being within the three most perceptually similar graphs.
Thus, we believe our WGL method is able to provide the expected layout results that are perceptually similar to the actual layout results.

When we analyze participants' responses for the predicted choices ($r_{\text{T}}=1$) for each layout separately, we find that the predicted choices with Spectral and Treemap layouts are ranked lower than with other layouts.
The common reasons given by participants for selecting choice graphs with Spectral layout were ``line shape'' and ``number of vertices''.
We notice that the Spectral layouts have many nodes that overlap each other.
Treemap, on the other hand, produces similar looking graphs due to its geometric constraints.
This observation was mirrored by many participants who said ``they all look similar'' for the choices with Treemap layout.
Common reasons for selecting choice graphs with Treemap layout were ``edge density'' and ``overall edge direction''.

It is interesting to see how people perceive certain structures as more important than others.
For instance, when we look at the responses on target graphs separately, we notice that target graph $G_{2331}$ has different response patterns.
Target graph $G_{2331}$ and its choice graph are shown in \autoref{fig:task}.
Participants' responses for $r_{\text{P}}=1$ are split between two choices, \autoref{fig:task}b and \autoref{fig:task}c.
The common reasons why the participants ranked \autoref{fig:task}c as the most similar to \autoref{fig:task}a were ``density'', ``shape'', and ``number of edges''.
On the other hand, the common reason why the participants ranked \autoref{fig:task}b as the most similar to \autoref{fig:task}a was ``the number of central nodes''.
Our method also chose \autoref{fig:task}c as the most similar graph because of the general structure matching the target graph, but ranked \autoref{fig:task}b as being second most similar $r_{\text{T}}=2$.
In the case of target graph $G_{2331}$, the number of nodes in the center held more value to some participants than the overall structure.

In our user study, only one similar graph was chosen and shown by our system per target graph. 
In a real system, the user would be given several similarly looking graphs, including isomorphic graphs. 
Thus, the real system would show more layouts closer to the actual one. 

\section{Related Work}
Only a handful of studies have used topological features for visualizing graphs.
Perhaps this is why there is a scarcity of studies applying machine learning techniques to the process of visualizing a graph \cite{Vieria15}.

Niggemann and Stein \cite{Niggemann00} introduced a learning method to find an optimal layout method for a clustered graph. 
The method constructs a handcrafted feature vector of a cluster from a number of graph measures, including the number of vertices, diameter, and maximum vertex degree. 
Then, it attempts to find an optimal layout for each cluster.
However, these features have been proved as not expressive enough to capture topological similarities in many graph kernel works \cite{Shervashidze12}.

Behrisch et al. \cite{Behrisch17} proposed a technique called Magnostics, where a graph is represented as a matrix view and image-based features are used to find similar matrices.
One of the challenges of a matrix view is the vertex ordering.
Depending on the ordering, even the same graph can be measured as a different graph from itself.
Graph kernels do not suffer from the same problem since they measure the similarity using only the topology of the graph.

Several techniques have used machine learning approaches to improve the quality of a graph layout \cite{Vieria15}.
Some of these techniques used evolutionary algorithms for learning user preferences with a human-in-the-loop assessment \cite{Masui94, Barbosa01, Bach12, Sponemann14}, while others have designed neural network algorithms to optimize a layout for certain aesthetic criteria \cite{Cimikowski96, Meyer98, Wang05}.
One major limitation of these techniques is that models learned from one graph are not usable in other graphs.
Since these techniques often require multiple computations of layouts and their aesthetic metrics, the learning process can be highly time-consuming.
These techniques can benefit from our approach by quickly showing the expected layout results and estimating the aesthetic metrics.

Many empirical studies have been conducted to understand the relation between topological characteristics of graphs and layout methods. 
The main idea is to find the ``best'' way to lay out given graphs \cite{Gibson12}. 
To achieve this, Archambault et al. \cite{Archambault07} introduced a layout method which first recursively detects the topological features of subgraphs, such as whether a subgraph is a tree, cluster, or complete graph. 
Then, each subgraph is laid out  using the suitable method according to its topological characteristics.
A drawback of this method is that the feature detectors are limited to five classes.
Our kernel can be utilized for heterogeneous feature detection with less computational cost.

A number of recent studies investigated sampling methods for large graph visualization.
Wu et al. \cite{Wu17} evaluated a number of graph sampling methods in terms of resulting visualization. 
They found that different visual features were preserved when different sampling strategies were used. 
Nguyen et al. \cite{Nguyen17} proposed a new family of quality metrics for large graphs based on a sampled graph.

\section{Conclusion}
We have developed a machine learning approach using graph kernels for the purpose of showing what a graph would look like in different layouts and their corresponding aesthetic metrics. 
We have also introduced a framework for designing graphlet kernels, which allows us to derive several new ones.
The estimations using our new kernels can be derived several orders of magnitude faster than computing the actual layouts and their aesthetic metrics.
Also, our kernels outperform state-of-the-art kernels in both accuracy and computation time.
The results of our user study show that the topological similarity computed with our kernel matches perceptual similarity assessed by human users.

In our work, we have only considered a subset of layout methods. 
A possible future direction is to include more layout methods with additional parameter settings for each method.
Mechanical Turk could be used to conduct such an experiment at scale.
Another possible future direction of this work is to introduce a new layout method which quickly predicts the actual layout instead of just showing the expected results of the input graph.
We hope this paper opens a new area of study into using machine learning
approaches for large graph visualization.

\acknowledgments{This research has been sponsored in part by the U.S. National Science Foundation through grants IIS-1320229 and IIS-1528203.}

\bibliographystyle{abbrv}

\bibliography{tex/references}
\end{document}